\documentclass[superscriptaddress,onecolumn,secnumarabic,
nobibnotes,aps,prd,showpacs,nofootinbib]{revtex4}
\usepackage{graphicx}
\usepackage{epsf}
\usepackage{bm}
\usepackage{amsmath}
\usepackage{amsfonts}
\usepackage{amssymb}
\usepackage{epstopdf}
\usepackage{natbib}
\usepackage{color}
\usepackage{color}

\providecommand{\U}[1]{\protect\rule{.1in}{.1in}}
\newcommand{\be}{\begin{equation}}
\newcommand{\ee}{\end{equation}}

\newcommand{\mincir}{\raise
-3.truept\hbox{\rlap{\hbox{$\sim$}}\raise4.truept\hbox{$<$}\ }}
\newcommand{\magcir}{\raise
-3.truept\hbox{\rlap{\hbox{$\sim$}}\raise4.truept\hbox{$>$}\ }}

\begin{document}

\title{Large-scale Stability and Astronomical Constraints for Coupled
Dark-Energy Models}
\author{Weiqiang Yang}
\email{d11102004@163.com}
\affiliation{Department of Physics, Liaoning Normal University, Dalian, 116029, P. R.
China}
\author{Supriya Pan}
\email{span@research.jdvu.ac.in}
\affiliation{Department of Physical Sciences, Indian Institute of Science Education and
Research, Kolkata, Mohanpur$-$741246, West Bengal, India}
\affiliation{Department of Mathematics, Raiganj Surendranath Mahavidyalaya, Sudarshanpur,
Raiganj, West Bengal 733134, India}
\author{John D. Barrow}
\email{jdb34@damtp.cam.ac.uk}
\affiliation{DAMTP, Centre for Mathematical Sciences, University of Cambridge,
Wilberforce Rd., Cambridge CB3 0WA, U.K.}

\begin{abstract}
We study large-scale inhomogeneous perturbations and instabilities of interacting dark energy (IDE) models. Past analysis of large-scale perturbative instabilities, has shown that we can only test IDE models with observational data when its parameter ranges are either $w_{x}\geq -1$ and $\xi \geq 0,$ or $w_{x}\leq -1~$ and $~\xi \leq 0$, where $w_{x}$ is the dark energy equation of state (EoS), and $\xi$ is a coupling parameter governing the strength and direction of the energy transfer. We show that by adding a factor $(1+w_{x})$ to the background energy transfer, the whole parameter space can be tested against all the data and thus, the instabilities in such interaction models can be removed. We test three classes of interaction model using the latest astronomical data from different sources. Precise constraints are found. Our analysis shows that a very small but non-zero deviation from pure $\Lambda$-cosmology is suggested by the observational data while the no-interaction scenario can be recovered at the 68.3\% confidence-level. In particular, for three IDE models, identified as IDE 1, IDE 2, and IDE 3, the 68.3\% CL constraints on the interaction coupling strengths are, $\xi= 0.0360_{-0.0360}^{+0.0091}$ (IDE 1), $\xi= 0.0433_{-0.0433}^{+0.0062}$ (IDE 2), $\xi= 0.1064_{-0.1064}^{+0.0437}$ (IDE 3). In addition, we find that the dark energy EoS tends towards the phantom region taking the 68.3\% CL constraints, $w_x= -1.0230_{-0.0257}^{+0.0329}$ (IDE 1), $w_x= -1.0247_{-0.0302}^{+0.0289}$ (IDE 2), and $w_x= -1.0275_{-0.0318}^{+0.0228}$ (IDE 3). However, the possibility of $w_{x}>-1$ is also not rejected by the astronomical data used here. Moreover, we find in all IDE models that, as the value of Hubble constant decreases, the behavior of the dark energy EoS shifts from phantom to quintessence type with its EoS very close to that a simple cosmological constant at the present time.
\end{abstract}

\maketitle



\section{Introduction}

The physics of the dark energy and the dark matter is still an open issue in
cosmology. The dark energy occupies about 68.5\% of the total energy density
of the universe today \cite{ref:Planck2015-3}, and is believed to accelerate
its observed expansion, but the physical nature, origin, and time evolution
of this dark energy remain unknown. On the other hand, the dark matter
sector (occupying almost 27.5\% of the total energy density of the
present-day universe) appears to be the principal gravitational influence on
the formation of large-scale structure in the universe and its existence is
supported by direct evidence from the spiral galaxy rotation curves and
cluster dynamics \cite{ref:Sofue2001}. At present, we have a many
dark-energy models \cite{cop, at} and, according to syntheses of all the
current observational data, $\Lambda $-cosmology appears to be the simplest
cosmological model that can explain the bulk of the evidence. However, the
unexplained numerical value of the cosmological constant, and the
coincidences between the present densities of the different dark and
luminous components of the universe, provoke us to search for new
cosmological scenarios in which the observed state of affairs is more
natural. In this work we will explore cosmologies where dark energy
interacts and exchanges energy with dark matter. \newline

Originally, the possibility that dark energy might interact with
dark matter was introduced to justify the very small value of the
cosmological constant by Wetterich \cite{Wetterich1, Wetterich2}. However,
when dynamical models were introduced as alternatives to a simple
(non-interacting) cosmological constant, it was found that interactions
between dark energy and dark matter might provide a simple explanation for
the cosmic coincidence problem \cite{Lip}.\ If one views this interaction
from the particle physics perspective, then it is natural that the two
fields should interact with each other non-gravitationally \cite%
{ref:Peebles2010}. Models of this type are known as interacting, or
coupled, dark energy models.\newline

The interacting dynamics is described by a new function $Q$, which
determines the form of the coupling between dark matter and dark energy 
via their conservation equations as $\nabla _{\nu }T_{c}^{\mu \nu
}=-Q$ and $\nabla _{\nu }T_{x}^{\mu \nu }=Q$, where $T_{c}^{\mu \nu }$, $%
T_{x}^{\mu \nu }$ are respectively identified as the energy-momentum tensors
for cold dark matter (CDM) and dark energy (DE). Consequently, one can
further identify $\rho _{c}$, $\rho _{x}$ to be the energy densities of CDM
and DE fluids, respectively. Until now, there have been many interacting
dark energy models based on different proposals for the form of energy
exchange term $Q$. A series of investigations have been performed using
observational data with interesting results
\cite{ref:Amendola2000,ref:Amendola2000-2,Billyard:2000bh,
ref:Zimdahl2001,Olivares:2005tb,
ref:Boehmer2008,ref:He2008,Quartin:2008px,Chimento:2009hj,
ref:Salvatelli2014,Pan:2012ki,Nunes:2016dlj,Kumar:2016zpg,
Marcondes:2016reb,Pan:2016ngu,Mukherjee:2016shl,Sharov:2017iue,Yang:2017yme,ref:Valiviita2008,ref:Majerotto2010,ref:Clemson2012, ref:Yang2014-uc,ref:Yang2014-ux,ref:Yang2014-dh,ref:Yang2016-four,Marttens:2016cba,Cai:2017yww}. 
Aside from the specific issue of dark matter-dark energy interactions, we
can also view the interaction, $Q$, as an energy exchange between any two
barotropic fluids, see \cite{Barrow:2006hia}.\newline

The interacting fluid models are generally well behaved when one only
considers their effects on the background evolution. However, the analysis
of inhomogeneous cosmological perturbations is essential to provide a fuller
picture of these models, to determine if they are stable or unstable
components of the large scale structure of the universe. For example, a
simple energy exchange term $Q\propto \rho _{c}$ leads to an instability in
the dark matter perturbations at early times since the curvature
perturbation blows up on super-Hubble scales \cite{ref:Valiviita2008}.%
 In order to derive a stable perturbation evolution, another simple
interaction term, $Q\propto \rho _{x}$ needs to be tested by the
observations with two intervals of possible dark energy equations of state: %
$w_{x}\leq -1$ and $w_{x}\geq -1$ \cite%
{ref:Clemson2012,ref:Yang2014-uc,ref:Yang2014-ux,ref:Yang2014-dh,ref:Yang2016-four}%
. Therefore, the principal motivation of this paper is to find a form of
energy transfer, $Q$, which could alleviate the perturbative instability. In
this way, we might test the full parameter space of dark energy
equations of state by the observations, allowing even for the
possibility of a 'phantom' equation of state. In this respect, large scale
structure information, such as redshift-space distortion (RSD) \cite%
{ref:RSD-Kaiser1987,ref:RSD-Hamilton1998,ref:fsigma83-Samushia2012,ref:fsigma8-DE-Song2009}
and weak gravitational lensing (WL) \cite%
{ref:Bartelmann2011,ref:Heymans2013,ref:Heymans2016}, provide an important
tools to break any degeneracy of cosmological models. This view has already
been confirmed by many investigations \cite%
{ref:Xu2013-HDE,ref:Xu2013-index,ref:Xu2013-DGP,ref:Yang2013-CASS,ref:Xu2014-vis1,ref:Xu2014-vis2,ref:Xu2015-fR,ref:Xu2015-MG,ref:Xu2015-phiCDM,ref:zhc2016-TDDE}%
. One conclusion from these studies was that joint measurements 
of the geometry and dynamical observations found that the
interaction rate, $Q$, was zero at about $1\sigma $ \cite%
{ref:Yang2014-uc,ref:Yang2014-ux,ref:Yang2014-dh,ref:Yang2016-four}.
Unfortunately, this conclusion has been drawn using the intervals $%
w_{x}\leq -1$ and $w_{x}\geq -1$ separately. If we could test the
interacting dark energy model with the full parameter space of $w_{x}$
against the observations, then a different conclusion might be found. 
This is an aim of this paper.

The paper is outlined as follows. In section \ref{sec-perturbations} we
describe the perturbation equations for the interacting dark-energy models.
Section \ref{sec-data} contains a brief description on the observational
data used in our analysis. In section \ref{sec-results} we discuss the main
observational results extracted from the interacting models in our study.
Finally, in section \ref{sec-summary} we conclude with a short summary.

\section{Background and perturbation evolution in coupled dark-energy models}

\label{sec-perturbations}

In this section we describe the dynamics of the coupled dark energy model at
both the background and perturbative levels. As usual, we consider a
spatially flat Friedmann-Lema\^{\i}tre-Robertson-Walker (FLRW) universe
characterized by the metric line element 
\begin{equation*}
ds^{2}=-dt^{2}+a^{2}(t)\left[ dr^{2}+r^{2}(d\theta ^{2}+\sin ^{2}\theta
d\phi ^{2})\right] ,
\end{equation*}%
where $a(t)$ is the expansion scale factor and $t$ is the comoving proper
time. The total energy density of the universe is $\rho _{t}=\rho _{c}+\rho
_{x}+\rho _{b}+\rho _{r}$, where we identify each $\rho _{i}$ as the energy
density of the $i$-th fluid component (the subscripts $c, $ $x$, $b$, $r$,
respectively, stand for cold dark matter, dark energy, baryons, and
radiation). The cold dark matter is pressureless, and we assume the dark
energy is barotropic. In order to neglect any kind of inflexible constraints
like a \textquotedblleft fifth force\textquotedblright , we assume that the
baryons and radiation are conserved separately; in other words, they follow
the usual conservation laws without any interaction. Now, in such a
spacetime, the modified conservation equations for cold dark matter and dark
energy are assumed to have the following forms, 
\begin{eqnarray}
\rho _{c}^{\prime }+3\mathcal{H}\rho _{c} &=&-aQ,  \label{cons1} \\
\rho _{x}^{\prime }+3\mathcal{H}(1+w_{x})\rho _{x} &=&aQ,  \label{cons2}
\end{eqnarray}%
where prime $^{\prime }$ denotes differentiation with respect to the
conformal time; $\mathcal{H}=a^{\prime }/a$ is the conformal Hubble
parameter; $w_{x}$ is the equation of state parameter of dark energy. The
positive energy exchange term shows that the energy transfer is from dark
matter to dark energy, and negative $Q$ denotes the opposite case. Further,
one can see that the conservation equations (\ref{cons1}) and (\ref{cons2})
can be rewritten by introducing effective equations of state for the dark
fluids as

\begin{eqnarray}
\rho _{c}^{\prime }+3\mathcal{H}\left( 1+w_{c}^{\mathtt{eff}}\right) \rho
_{c} &=&0,  \notag \\
\rho _{x}^{\prime }+3\mathcal{H}\left( 1+w_{x}^{\mathtt{eff}}\right) \rho
_{x} &=&0,  \notag
\end{eqnarray}%
where $w_{c}^{\mathtt{eff}}$, $w_{x}^{\mathtt{eff}}$ are defined as the
effective equation of state parameters for CDM and dark energy with 
\begin{eqnarray}
w_{c}^{\mathtt{eff}} &=&\frac{aQ}{3\mathcal{H}\rho _{c}},  \notag \\
w_{x}^{\mathtt{eff}} &=&w_{x}-\frac{aQ}{3\mathcal{H}\rho _{x}}.  \notag
\end{eqnarray}%
We note that the effective equation of state parameter for CDM could be
nonzero while the effective equation of state for dark energy offers several
possibilities depending on the strength of the interaction rate, $Q$. In
particular, the direction of energy transfer controls the nature of an
effective dark energy (`phantom' or `quintessence' or an `equivalent
cosmological constant' scenario) fluid through the quantity $w_{x}^{\mathtt{%
eff}}$. Finally, the Friedmann equation is

\begin{equation*}
\mathcal{H}^{2}=\frac{8\pi G}{3}a^{2}\left( \rho _{c}+\rho _{x}+\rho
_{b}+\rho _{r}\right) ,
\end{equation*}%
which constrains the dynamics of the universe. Thus, the system of equations
(\ref{cons1}), (\ref{cons2}) together with the Friedmann equation determines
the entire dynamics of the universe, once the energy transfer rate $Q$ is
specified.

We shall now discuss the linear perturbations for the interacting models
that we introduce here. The metric that determines the most general scalar
mode perturbation is given by \cite%
{ref:Ma1995,ref:Mukhanov1992,ref:Malik2009} 
\begin{equation*}
ds^{2}=a^{2}(\tau )\Bigl[-(1+2\phi )d\tau ^{2}+2\partial _{i}Bd\tau dx^{i}+%
\Bigl((1-2\psi )\delta _{ij}+2\partial _{i}\partial _{j}E\Bigr)dx^{i}dx^{j}%
\Bigr],
\end{equation*}%
where the quantities $\phi $, $B$, $\psi $ and $E$, respectively stand for
the gauge-dependent scalar perturbations and $\tau $ is the conformal time.
Now for any fluid subscripted by `$A$', its energy-momentum conservation
equations can be calculated and are \cite%
{ref:Majerotto2010,ref:Valiviita2008,ref:Clemson2012}, 
\begin{equation*}
\nabla _{\nu }T_{A}^{\mu \nu }=Q_{A}^{\mu },~~~~\sum\limits_{\mathrm{A}}{%
Q_{A}^{\mu }}=0,
\end{equation*}%
where one has $Q_{A}^{\mu }=(Q_{A}+\delta Q_{A})u^{\mu }+a^{-1}(0,\partial
^{i}f_{A})$ relative to the four-velocity $u^{\mu }$, \cite%
{ref:Majerotto2010,ref:Valiviita2008,ref:Clemson2012}. We specialize the
momentum transfer potential to be the simplest physical choice, which is
zero in the rest frame of the dark matter \cite%
{ref:Valiviita2008,ref:Koyama2009,ref:Clemson2012}. Hence, the momentum
transfer potential becomes $k^{2}f_{A}=Q_{A}(\theta -\theta _{c})$. We
define the pressure perturbation by $\delta p_{A}=c_{sA}^{2}\delta \rho
_{A}+(c_{sA}^{2}-c_{aA}^{2})\rho _{A}^{\prime }(v_{A}+B)$ \cite%
{ref:Kodama1984,ref:Hu1998,ref:Valiviita2008}, where $c_{aA}^{2}=p_{A}^{%
\prime }/\rho _{A}^{\prime }=w_{x}+w_{x}^{\prime }/(\rho _{A}^{\prime }/\rho
_{A})$, is the physical sound speed of the fluid `$A$' in the rest frame. If
we further define the density contrast by $\delta _{A}=\delta \rho _{A}/\rho
_{A}$ and consider $\pi _{A}=0$, then in the synchronous gauge,
equivalently, $\phi =B=0$, $\psi =\eta $, and $k^{2}E=-h/2-3\eta $, the
general evolution equations for the density perturbation (i.e. the
continuity equation) and the velocity perturbation (Euler equation)
equations for dark energy and dark matter respectively, become 
\begin{eqnarray}
\delta _{x}^{\prime } &=&-(1+w_{x})\left( \theta _{x}+\frac{h^{\prime }}{2}%
\right) -3\mathcal{H}(c_{sx}^{2}-w_{x})\left[ \delta _{x}+3\mathcal{H}%
(1+w_{x})\frac{\theta _{x}}{k^{2}}\right] -3\mathcal{H}w_{x}^{\prime }\frac{%
\theta _{x}}{k^{2}}  \notag \\
&+&\frac{aQ}{\rho _{x}}\left[ -\delta _{x}+\frac{\delta Q}{Q}+3\mathcal{H}%
(c_{sx}^{2}-w_{x})\frac{\theta _{x}}{k^{2}}\right] , \\
\theta _{x}^{\prime } &=&-\mathcal{H}(1-3c_{sx}^{2})\theta _{x}+\frac{%
c_{sx}^{2}}{(1+w_{x})}k^{2}\delta _{x}+\frac{aQ}{\rho _{x}}\left[ \frac{%
\theta _{c}-(1+c_{sx}^{2})\theta _{x}}{1+w_{x}}\right] , \\
\delta _{c}^{\prime } &=&-\left( \theta _{c}+\frac{h^{\prime }}{2}\right) +%
\frac{aQ}{\rho _{c}}\left( \delta _{c}-\frac{\delta Q}{Q}\right) , \\
\theta _{c}^{\prime } &=&-\mathcal{H}\theta _{c},  \label{eq:perturbation}
\end{eqnarray}%
where the term $\delta Q/Q$ includes the perturbation term for the Hubble
expansion rate $\delta H$ (we note that $\mathcal{H}=aH$). From the
perturbation of the Hubble expansion rate, $\delta H$, one could obtain the
gauge invariant equations for the coupled dark sector \cite{ref:Gavela2009}.
Thus, we consider the perturbation of the Hubble expansion rate since the
total expansion rate would include two parts: background and perturbation.
In the light of the analysis of the contribution from the perturbation of
the expansion rate in ref. \cite{ref:Gavela2009}, it is chosen to be
associated with the volume expansion of the total fluid, i.e., $\delta
H/H=(\theta +h^{\prime }/2)/(3\mathcal{H})$.

The energy transfer may change the history of the universe. In most of the
cases, interacting models are reliable when their background evolution is
considered. However, it is also very important to take care of the
cosmological perturbations in order to ensure the stability of the
cosmological models under consideration. The Hubble rate is assumed to be
the average expansion rate in $Q$. One should treat $H$ as a local variable
so as to include the perturbation term $\delta H$. Thus, we can consistently
obtain the gauge-invariant perturbation equations \cite{ref:Gavela2010}.

In the following we shall discuss the stability and instability issues
associated with the current interacting models. The large-scale instability
arises from the pressure perturbation of dark energy \cite{ref:Valiviita2008}%
. The pressure perturbation includes the adiabatic pressure perturbation and
the intrinsic non-adiabatic pressure perturbation. For the interacting dark
energy models, the non-adiabatic part might grow fast at early times due to
the energy transfer and this leads to rapid growth of the curvature
perturbation on the large scales. For example, as mentioned above, the
simple energy exchange term $Q\propto \rho _{c}$ leads to an instability in
the dark matter perturbations at early times since the curvature
perturbation blows up on super-Hubble scales \cite{ref:Valiviita2008}.
Subsequently, another interaction model $Q=3H\xi \rho _{c}\rho _{x}/(\rho
_{c}+\rho _{x})$ was suggested in ref. \cite{Li:2013bya}, where it was shown
that this form of $Q$ for the energy transfer could avoid the large-scale
instability during the early expansion of the universe.

The pressure perturbation for the coupled dark energy models is given by 
\cite{ref:Hu1998,ref:Gavela2009} 
\begin{eqnarray}
\delta p_{x} &=&c_{sx}^{2}\delta \rho _{x}-(c_{sx}^{2}-c_{ax}^{2})\rho
_{x}^{\prime }\frac{\theta _{x}}{k^{2}}~,  \notag \\
&=&c_{sx}^{2}\delta \rho _{x}+3\mathcal{H}\rho
_{x}(1+w_{x})(c_{sx}^{2}-c_{ax}^{2})\left[ 1-\frac{aQ}{3\mathcal{H}\rho
_{x}(1+w_{x})}\right] \frac{\theta _{x}}{k^{2}}~,  \notag \\
&=&c_{sx}^{2}\delta \rho _{x}+3\mathcal{H}\rho
_{x}(1+w_{x})(c_{sx}^{2}-c_{ax}^{2})(1+d)\frac{\theta _{x}}{k^{2}}~.
\label{eq:deltap}
\end{eqnarray}%
Now, one could judge the stability condition of the perturbations via the
`doom factor' \cite{ref:Gavela2009}, defined as 
\begin{equation*}
d\equiv -aQ/[3\mathcal{H}\rho _{x}(1+w_{x})],
\end{equation*}%
using the pressure perturbation of dark energy. Thus, stability can be
realized when $d\leq 0$ \cite{ref:Gavela2009,ref:Clemson2012}. It means that
for the usual interaction rates in the literature,  
$Q = 3 H \xi \bar{Q}$ (with $\bar{Q} > 0$),  
the perturbation stability requires the conditions $\xi \geq 0~\&~(1+w_{x})>0$
or $\xi \leq 0~\&~(1+w_{x})<0$. Following this, interaction term $Q=3H\xi
\rho _{x}$ needs to be tested against the observations with two intervals
for dark-energy equation of state $w_{x}\leq -1$ and $w_{x}\geq -1$ \cite%
{ref:Clemson2012,ref:Yang2014-uc,ref:Yang2014-ux,ref:Yang2014-dh,ref:Yang2016-four}%
. We note that $w_{x}=-1$ is the limiting case, see \cite{ref:Clemson2012}
for details. Now looking at the pressure perturbations in eqn. (\ref%
{eq:deltap}), it is worth to note that the interaction functions with $%
(1+w_{x})$ could release the prior of DE equation of state (EoS) which is a
very interesting property because the prior on the dark energy equation of
state plays a crucial role in the statistical analysis. Thus, here we will
assume a phenomenological energy transfer which includes the factor $%
(1+w_{x})$ explicitly, for example of the form $Q=3H\xi (1+w_{x})\rho _{x}$, 
$Q=3H\xi (1+w_{x})\rho _{c}\rho _{x}/(\rho _{c}+\rho _{x})$, $Q=3H\xi
(1+w_{x})\rho _{x}^{\alpha }\rho _{c}^{\beta }$, or the general form $%
Q=3H\xi (1+w_{x})\rho _{c}^{\alpha }\rho _{x}^{\gamma }(\rho _{c}+\rho
_{x})^{\beta }$, where $w_{x}$ might be constant or time-dependent. Thus, we
can define the doom factor for the coupled model 
\begin{equation*}
d\equiv -\frac{aQ}{3\mathcal{H}\rho _{x}(1+w_{x})}=-\xi \rho _{c}^{\alpha
}\rho _{x}^{\gamma -1}(\rho _{c}+\rho _{x})^{\beta }.
\end{equation*}%
Now, it is easy to see that in order to have the stable perturbations, i.e. $%
d\leq 0,$ the coupling parameter should satisfy the relation $\xi \geq 0$.
That means there is no need to test the interaction models for two intervals
of dark energy equation of state, namely, $w_{x}\leq -1$ and $w_{x}\geq -1$ 
\cite{ref:Yang2014-uc,ref:Yang2014-ux,ref:Yang2014-dh,ref:Yang2016-four};
rather, we could just constrain the full parameter space of $w_{x}$ using
the observational data. Thus, with the simple constraint on the coupling
parameter that $\xi \geq 0$, we can alleviate the large-scale perturbation
instabilities in the coupled dark-energy models - this is the novelty of the
present work. In this way, we can explore the possibility of a phantom
dark-energy equation of state. It should be noted that, for some suitable
time-varying dark energy equations of state, such as the Chevallier-Polarski
and Linder (CPL) parametrization \cite{cpl1,cpl2}, the perturbation
instability could also be alleviated \cite{ref:Majerotto2010}. However, we
note that the proposed general interaction model $Q=3H\xi (1+w_{x})\rho
_{c}^{\alpha }\rho _{x}^{\gamma }(\rho _{c}+\rho _{x})^{\beta } $ can be
viewed as $Q=3H\bar{\xi}\rho _{c}^{\alpha }\rho _{x}^{\gamma }(\rho
_{c}+\rho _{x})^{\beta }$ using a simple transformation $\xi \rightarrow 
\bar{\xi}=\xi (1+w_{x})$. Now, we observe that, if one allows the dark
energy equation of state to run beyond the cosmological constant limit, i.e. 
$w_{x}\leq -1$, then considering the stability condition $d\leq 0$, the
model could produce stable perturbations on the large scales for $\bar{\xi}%
\leq 0$. This is an alternative route to produce the stable perturbations
from interaction models for the large scale structure of the universe \cite%
{ref:Clemson2012, ref:Yang2014-uc, ref:Yang2014-ux, ref:Yang2014-dh,
ref:Yang2016-four} without introducing the factor $(1+w_{x})$ explicitly
outside the interaction rate.

Next, we recall the general interaction model which recovers the three
interactions used in our study above. Since this general interaction assumes
the expression

\begin{equation*}
Q=3H\xi (1+w_{x})\rho _{c}^{\alpha }\rho _{x}^{\gamma }(\rho _{c}+\rho
_{x})^{\beta },
\end{equation*}%
where the exponents $(\alpha ,\beta ,\gamma )\in \mathbb{R}^{3}$ must
satisfy $\alpha +\beta +\gamma =1,$ so that the dimension of $Q$ is in
accord with the background energy-momentum conservation equation, then,
using the relation $\gamma =1-\alpha -\beta $, we may rewrite $Q$ as $%
Q=3H\xi (1+w_{x})\rho _{c}^{\alpha }\rho _{x}^{1-\alpha -\beta }(\rho
_{c}+\rho _{x})^{\beta }$. Now, for this interaction the variation $\delta Q$
reads 
\begin{equation*}
\delta Q=Q\left[ \alpha \delta _{c}+(1-\alpha -\beta )\delta _{x}+\beta 
\frac{\rho _{c}\delta _{c}+\rho _{x}\delta _{x}}{\rho _{c}+\rho _{x}}+\frac{%
\theta +h^{\prime }/2}{3\mathcal{H}}\right] ,
\end{equation*}%
and consequently, the density and velocity perturbation equations for dark
energy and dark matter for this $Q$ become 
\begin{eqnarray}
\delta _{x}^{\prime } &=&-(1+w_{x})\left( \theta _{x}+\frac{h^{\prime }}{2}%
\right) -3\mathcal{H}(c_{sx}^{2}-w_{x})\left[ \delta _{x}+3\mathcal{H}%
(1+w_{x})\frac{\theta _{x}}{k^{2}}\right] -3\mathcal{H}w_{x}^{\prime }\frac{%
\theta _{x}}{k^{2}}  \notag \\
&+&3\mathcal{H}\xi (1+w_{x})\rho _{c}^{\alpha }\rho _{x}^{-\alpha -\beta
}(\rho _{c}+\rho _{x})^{\beta }\left[ \alpha \delta _{c}-(\alpha +\beta
)\delta _{x}+\beta \frac{\rho _{c}\delta _{c}+\rho _{x}\delta _{x}}{\rho
_{c}+\rho _{x}}+\frac{\theta +h^{\prime }/2}{3\mathcal{H}}+3\mathcal{H}%
(c_{sx}^{2}-w_{x})\frac{\theta _{x}}{k^{2}}\right] , \\
\theta _{x}^{\prime } &=&-\mathcal{H}(1-3c_{sx}^{2})\theta _{x}+\frac{%
c_{sx}^{2}}{(1+w_{x})}k^{2}\delta _{x}+3\mathcal{H}\xi \rho _{c}^{\alpha
}\rho _{x}^{-\alpha -\beta }(\rho _{c}+\rho _{x})^{\beta }\left[ \theta
_{c}-(1+c_{sx}^{2})\theta _{x}\right] , \\
\delta _{c}^{\prime } &=&-\left( \theta _{c}+\frac{h^{\prime }}{2}\right) 
\notag \\
&+&3\mathcal{H}\xi (1+w_{x})\rho _{c}^{\alpha -1}\rho _{x}^{1-\alpha -\beta
}(\rho _{c}+\rho _{x})^{\beta }\left[ (1-\alpha )\delta _{c}-(1-\alpha
-\beta )\delta _{x}-\beta \frac{\rho _{c}\delta _{c}+\rho _{x}\delta _{x}}{%
\rho _{c}+\rho _{x}}-\frac{\theta +h^{\prime }/2}{3\mathcal{H}}\right] , \\
\theta _{c}^{\prime } &=&-\mathcal{H}\theta _{c},
\label{eq:perturbation-general}
\end{eqnarray}%
where $\alpha \leq 0$ or $\beta \leq 0$ are required for the perturbation
evolution to be stable at early times, according to the analysis of
large-scale instability \cite{ref:Valiviita2008}. These perturbation
equations of dark energy and dark matter include a many coupled dark-energy
models. For example, if $\beta =0$, the stability requirement $\alpha \leq 0$
favors the coupling $Q=3H\xi (1+w_{x})\rho _{c}^{\alpha }\rho _{x}^{1-\alpha
}$. For $\alpha =-1$ and $\beta =0$, we get the coupling $Q=3H\xi
(1+w_{x})\rho _{x}^{2}/\rho _{c}$. When $\alpha =1$ and $\beta =-1$, we have 
$Q=3H\xi (1+w_{x})\rho _{c}\rho _{x}/(\rho _{c}+\rho _{x})$. Further, for $%
\alpha =\beta =0$, we could obtain the simplest energy transfer, with $%
Q=3H\xi (1+w_{x})\rho _{x}$. We note that the explicitly appearance of the
Hubble factor $H$, in the interaction function is in general not necessary
in spatially-flat universes. However, its appearance helps us to write the
conservation equations with respect to the lapse function or the scale
factor of the FLRW universe\footnote{%
The conservation equations (\ref{cons1}) and (\ref{cons2}) can respectively
be rewritten as $\rho _{m}^{\prime }+3\rho _{m}=-\bar{Q}$ and $\rho
_{x}^{\prime }+3(1+w_{x})\rho _{x}=\bar{Q}$, where $\bar{Q}=Q/H$ and the
prime is taken with respect to the lapse function $N=\ln a$.}. Moreover, the
volume factor `$3$' has no physical meaning, this is just for simplicity
without any loss of generality. We note that the perturbation equations are
valid when the dark-energy equation of state is time dependent, such as CPL 
\cite{cpl1, cpl2} and similar. Now, for some particular choices of $\alpha $%
, $\beta $, we will test three interacting dark energy models against the
observational data sets when the dark energy equation of state is assumed to
be constant.\newline


We consider first the simplest interacting dark energy model (labelled IDE
1), with $\alpha =0$ and $\beta =0$. The coupling $Q$ thus becomes 
\begin{equation*}
Q=3H\xi (1+w_{x})\rho _{x}.
\end{equation*}%
For this model, following \cite{ref:Gavela2010}, we calculate that $\delta
Q=Q[\delta _{x}+(\theta +h^{\prime }/2)/(3\mathcal{H})]$. Thus, the
perturbation equations for the dark energy and dark matter become 
\begin{eqnarray}
\delta _{x}^{\prime } &=&-(1+w_{x})\left( \theta _{x}+\frac{h^{\prime }}{2}%
\right) -3\mathcal{H}(c_{sx}^{2}-w_{x})\left[ \delta _{x}+3\mathcal{H}%
(1+w_{x})\frac{\theta _{x}}{k^{2}}\right]  \notag \\
&+&3\mathcal{H}\xi (1+w_{x})\left[ \frac{\theta +h^{\prime }/2}{3\mathcal{H}}%
+3\mathcal{H}(c_{sx}^{2}-w_{x})\frac{\theta _{x}}{k^{2}}\right] , \\
\theta _{x}^{\prime } &=&-\mathcal{H}(1-3c_{sx}^{2})\theta _{x}+\frac{%
c_{sx}^{2}}{(1+w_{x})}k^{2}\delta _{x}+3\mathcal{H}\xi \left[ \theta
_{c}-(1+c_{sx}^{2})\theta _{x}\right] , \\
\delta _{c}^{\prime } &=&-\left( \theta _{c}+\frac{h^{\prime }}{2}\right) +3%
\mathcal{H}\xi (1+w_{x})\frac{\rho _{x}}{\rho _{c}}\left( \delta _{c}-\delta
_{x}-\frac{\theta +h^{\prime }/2}{3\mathcal{H}}\right) , \\
\theta _{c}^{\prime } &=&-\mathcal{H}\theta _{c}.  \label{eq:perturbation1}
\end{eqnarray}


\vspace{0.2cm}

Next we consider the second interaction model (IDE 2) for the specific
values of the parameters $\alpha =1$ and $\beta =-1$. The coupling for such
choice becomes 
\begin{equation*}
Q=3H\xi (1+w_{x})\frac{\rho _{c}\rho _{x}}{(\rho _{c}+\rho _{x})}.
\end{equation*}%
Consequently, we has $\delta Q=Q[\delta _{c}+\delta _{x}-(\rho _{c}\delta
_{c}+\rho _{x}\delta _{x})/(\rho _{c}+\rho _{x})+(\theta +h^{\prime }/2)/(3%
\mathcal{H})]$, and similarly the perturbation equations of dark energy and
dark matter follow as, 
\begin{eqnarray}
\delta _{x}^{\prime } &=&-(1+w_{x})\left( \theta _{x}+\frac{h^{\prime }}{2}%
\right) -3\mathcal{H}(c_{sx}^{2}-w_{x})\left[ \delta _{x}+3\mathcal{H}%
(1+w_{x})\frac{\theta _{x}}{k^{2}}\right]  \notag \\
&+&3\mathcal{H}\xi (1+w_{x})\frac{\rho _{c}}{\rho _{c}+\rho _{x}}\left[
\delta _{c}-\frac{\rho _{c}\delta _{c}+\rho _{x}\delta _{x}}{\rho _{c}+\rho
_{x}}+\frac{\theta +h^{\prime }/2}{3\mathcal{H}}+3\mathcal{H}%
(c_{sx}^{2}-w_{x})\frac{\theta _{x}}{k^{2}}\right] , \\
\theta _{x}^{\prime } &=&-\mathcal{H}(1-3c_{sx}^{2})\theta _{x}+\frac{%
c_{sx}^{2}}{(1+w_{x})}k^{2}\delta _{x}+3\mathcal{H}\xi \frac{\rho _{c}}{\rho
_{c}+\rho _{x}}\left[ \theta _{c}-(1+c_{sx}^{2})\theta _{x}\right] , \\
\delta _{c}^{\prime } &=&-\left( \theta _{c}+\frac{h^{\prime }}{2}\right) +3%
\mathcal{H}\xi (1+w_{x})\frac{\rho _{x}}{\rho _{c}+\rho _{x}}\left[ -\delta
_{x}+\frac{\rho _{c}\delta _{c}+\rho _{x}\delta _{x}}{\rho _{c}+\rho _{x}}-%
\frac{\theta +h^{\prime }/2}{3\mathcal{H}}\right] , \\
\theta _{c}^{\prime } &=&-\mathcal{H}\theta _{c}.  \label{eq:perturbation2}
\end{eqnarray}

\vspace{0.2cm}


Finally, we consider the third interaction model (IDE 3), with the choices $%
\alpha =-1$, $\beta =0$, and this leads to the coupling

\begin{equation*}
Q=3H\xi (1+w_{x})\frac{\rho _{x}^{2}}{\rho _{c}},
\end{equation*}%
which gives rise to the variation $\delta Q=Q[(-\delta _{c}+2\delta
_{x}+(\theta +h^{\prime }/2)/(3\mathcal{H})]$, and consequently, it is
possible to find the perturbation equations of dark energy and dark matter
respectively as 
\begin{eqnarray}
\delta _{x}^{\prime } &=&-(1+w_{x})\left( \theta _{x}+\frac{h^{\prime }}{2}%
\right) -3\mathcal{H}(c_{sx}^{2}-w_{x})\left[ \delta _{x}+3\mathcal{H}%
(1+w_{x})\frac{\theta _{x}}{k^{2}}\right]  \notag \\
&+&3\mathcal{H}\xi (1+w_{x})\frac{\rho _{x}}{\rho _{c}}\left[ -\delta
_{c}+2\delta _{x}+\frac{\theta +h^{\prime }/2}{3\mathcal{H}}+3\mathcal{H}%
(c_{sx}^{2}-w_{x})\frac{\theta _{x}}{k^{2}}\right] , \\
\theta _{x}^{\prime } &=&-\mathcal{H}(1-3c_{sx}^{2})\theta _{x}+\frac{%
c_{sx}^{2}}{(1+w_{x})}k^{2}\delta _{x}+3\mathcal{H}\xi \frac{\rho _{x}}{\rho
_{c}}\left[ \theta _{c}-(1+c_{sx}^{2})\theta _{x}\right] , \\
\delta _{c}^{\prime } &=&-\left( \theta _{c}+\frac{h^{\prime }}{2}\right) +3%
\mathcal{H}\xi (1+w_{x})\frac{\rho _{x}^{2}}{\rho _{c}^{2}}\left[ 2\delta
_{c}-2\delta _{x}-\frac{\theta +h^{\prime }/2}{3\mathcal{H}}\right] , \\
\theta _{c}^{\prime } &=&-\mathcal{H}\theta _{c}.  \label{eq:perturbation3}
\end{eqnarray}

We shall analyze these three interaction models using the latest
observational data and discuss their large scale stability.

\section{Observational data sets}

\label{sec-data}

To constrain the three interacting models (IDE 1-3) we use observational
data from different astronomical sources, as follows:

\begin{enumerate}
\item \textit{Cosmic microwave background observations (CMB):} We use CMB
data from the Planck 2015 measurements \cite{ref:Planck2015-1,
ref:Planck2015-2}, where we combine the full likelihoods $C_{l}^{TT}$, $%
C_{l}^{EE}$, $C_{l}^{TE}$ with low$-l$ polarization $%
C_{l}^{TE}+C_{l}^{EE}+C_{l}^{BB}$, which is notationally the same as the
\textquotedblleft PlanckTT, TE, EE + lowP\textquotedblright\ of ref. \cite%
{ref:Planck2015-2}.

\item \textit{Supernovae Type Ia:} Supernovae Type Ia are the first
geometric sample to infer the accelerating phase of the universe and so far
serve as one of the best samples to analyze any dark-energy model. In this
work we use the most latest SNIa sample known as Joint Light Curve Analysis
(JLA) samples \cite{Betoule:2014frx} comprising 740 data points in the
redshift range $0.01\leq z\leq 1.30$.

\item \textit{Baryon acoustic oscillation (BAO) distance measurements:} For
this data set, we use four BAO points: the 6dF Galaxy Survey (6dFGS)
measurement at $z_{\emph{\emph{eff}}}=0.106$ \cite{ref:BAO1-Beutler2011},
the Main Galaxy Sample of Data Release 7 of Sloan Digital Sky Survey
(SDSS-MGS) at $z_{\emph{\emph{eff}}}=0.15$ \cite{ref:BAO2-Ross2015}, and the
CMASS and LOWZ samples from the latest Data Release 12 (DR12) of the Baryon
Oscillation Spectroscopic Survey (BOSS) at $z_{\mathrm{eff}}=0.57$ \cite%
{ref:BAO3-Gil-Marn2015} and $z_{\mathrm{eff}}=0.32$ \cite%
{ref:BAO3-Gil-Marn2015}.

\item \textit{Redshift space distortion (RSD):} We employ two RSD
measurements, which include the CMASS sample with an effective redshift of $%
z_{\mathrm{eff}}=0.57$ and the LOWZ sample with an effective redshift of $z_{%
\mathrm{eff}}=0.32$~\cite{ref:RSD}.

\item \textit{Weak lensing (WL):} We use the weak gravitational lensing data
from Canada-France-Hawaii Telescope Lensing Survey (CFHTLenS) \cite%
{Heymans:2013fya,Asgari:2016xuw}.

\item \textit{Cosmic Chronometers (CC):} The Hubble parameter measurements
from most old and passively evolving galaxies, known as cosmic chronometers
(CC) have been considered to be potential candidates to probe the nature of
dark energy due to their model-independent measurements. For a detailed
description on how one can measure the Hubble parameter values at different
redshifts through this CC approach, and its usefulness, we refer to \cite%
{Moresco:2016mzx}. Here, we use 30 measurements of the Hubble parameter at
different redshifts within the range $0<z<2$.

\item \textit{Local value of the Hubble constant ($H_{0}$):} We include the
local value of the Hubble parameter which yields $H_{0}=73.24\pm 1.74$
km/s/Mpc with 2.4\% precision \cite{Riess:2016jrr}.
\end{enumerate}

\section{Results of the analysis}

\label{sec-results}

For the three interacting dark energy models above, we consider the
following eight-dimensional parameter space (see \cite{gen}) 
\begin{equation*}
P\equiv \{\Omega _{b}h^{2},\Omega _{c}h^{2},\Theta _{S},\tau ,w_{x},\xi
,n_{s},\log [10^{10}A_{S}]\},
\end{equation*}%
where $\Omega _{b}h^{2}$ and $\Omega _{c}h^{2}$ stand for the density of
baryons and the dark matter, respectively; $\Theta _{S}=100\theta _{MC}$ is
the ratio of sound horizon to the angular diameter distance; $\tau $ is the
optical depth; $w_{x}$ is the equation of state parameter of dark energy; $%
\xi $ is the coupling parameter; $n_{s}$ is the scalar spectral index; $%
A_{s} $ represents the amplitude of the initial power spectrum. The priors
of the basic model parameters are shown in the second column of Table \ref%
{tab:results}. The recent value of the Hubble constant $H_{0}=73.24\pm 1.74$
km/s/Mpc \cite{Riess:2016jrr} is used as a prior. Here, we note that the
sound speed of dark energy perturbations, $c_{sx\text{ }},$ plays an
important role in the large scale dynamics. For stable perturbations of dark
energy, one must have $c_{sx}^{2}>0$. Since we have assumed a constant
equation of state parameter for the dark energy, if dark energy is an
adiabatic fluid, then one can see that, $c_{sx}^{2}=c_{ax}^{2}=w_{x}<0$.
This means that the sound speed of dark energy perturbations becomes
imaginary, and consequently this leads to instabilities in the dark energy
evolution. Here, we assume $c_{sx}^{2}=1$, the sound speed for quintessence
following the earlier studies in refs \cite%
{ref:Valiviita2008,ref:Majerotto2010,ref:Clemson2012}. In fact, with the
assumption of $c_{sx}^{2}=1,$ or close to $1$, the dark energy does not
cluster on the sub-Hubble scale. The dark-matter velocity perturbation
equation is the same as in the uncoupled case, so we can consistently set $%
\theta _{c}=0$ \cite{ref:Valiviita2008}, since there is no momentum transfer
in the rest frame of dark matter. Here, in order to study the effects of the
interaction rate on the angular CMB power spectra, we modified the publicly
available CAMB package \cite{ref:camb}, which is included in CosmoMC \cite%
{ref:cosmomc-Lewis2002}, to calculate the anisotropic power spectrum of the
CMB.\newline

This allows us to analyze the results of global fitting for the three
different interaction models, namely, IDE 1: $Q=3H\xi (1+w_{x})\rho _{x}$,
IDE 2: $Q=3H\xi (1+w_{x})\rho _{c}\rho _{x}/(\rho _{c}+\rho _{x})$, and IDE
3: $Q=3H\xi (1+w_{x})\rho _{x}^{2}/\rho _{c}$.

The Table \ref{tab:results} summarizes the main observational results
extracted from all three interacting dark energy models using the combined
analysis CMB $+$ BAO $+$ JLA $+$ RSD $+$ WL $+$ CC $+$ $H_{0}$. In the following,
we describe the behaviour of each interacting fluid model in detail. \newline


\begin{center}
\begin{table}[tbp]
\begin{tabular}{cccccccc}
\hline\hline
Parameters & Priors & IDE 1 & Best fit & IDE 2 & Best fit & IDE 3 & Best fit
\\ \hline
$\Omega_bh^2$ & $[0.005,0.1]$ & $0.0223_{- 0.0001- 0.0003}^{+ 0.0001+
0.0003} $ & $0.0222$ & $0.0223_{- 0.0002- 0.0003}^{+ 0.0001+ 0.0003}$ & $%
0.0222$ & $0.0223_{- 0.0001- 0.0003}^{+ 0.0002+ 0.0003}$ & $0.0222$ \\ 
$\Omega_ch^2$ & $[0.01,0.99]$ & $0.1183_{- 0.0014- 0.0029}^{+ 0.0014+
0.0030} $ & $0.1185$ & $0.1182_{- 0.0012- 0.0027}^{+ 0.0013+ 0.0025}$ & $%
0.1186$ & $0.1194_{- 0.0023- 0.0047}^{+ 0.0022+ 0.0048}$ & $0.1180$ \\ 
$100\theta_{MC}$ & $[0.5,10]$ & $1.0406_{- 0.0003- 0.0006}^{+ 0.0003+
0.0006} $ & $1.0403$ & $1.0406_{- 0.0003- 0.0007}^{+ 0.0004+ 0.0006}$ & $%
1.0408$ & $1.0406_{- 0.0003- 0.0006}^{+ 0.0003+ 0.0006}$ & $1.0406$ \\ 
$\tau$ & $[0.01,0.8]$ & $0.0663_{- 0.0162- 0.0319}^{+ 0.0161+ 0.0315}$ & $%
0.0762$ & $0.0662_{- 0.0178- 0.0298}^{+ 0.0154+ 0.0318}$ & $0.0514$ & $%
0.0682_{- 0.0170- 0.0316}^{+ 0.0168+ 0.0317}$ & $0.0699$ \\ 
$n_s$ & $[0.5,1.5]$ & $0.9760_{- 0.0038- 0.0070}^{+ 0.0036+ 0.0071}$ & $%
0.9778$ & $0.9763_{- 0.0044- 0.0087}^{+ 0.0044+ 0.0085}$ & $0.9717$ & $%
0.9762_{- 0.0042- 0.0074}^{+ 0.0038+ 0.0079}$ & $0.9794$ \\ 
$\mathrm{ln}(10^{10}A_s)$ & $[2.4,4]$ & $3.0722_{- 0.0288- 0.0616}^{+
0.0311+ 0.0605}$ & $3.0945$ & $3.0714_{- 0.0341- 0.0607}^{+ 0.0302+ 0.0622}$
& $3.0414$ & $3.0747_{- 0.0332- 0.0623}^{+ 0.0333+ 0.0624}$ & $3.0747$ \\ 
$w_x$ & $[-2,0]$ & $-1.0230_{- 0.0257- 0.0603}^{+ 0.0329+ 0.0527}$ & $%
-1.0210 $ & $-1.0247_{- 0.0302- 0.0841}^{+ 0.0289+ 0.0895}$ & $-1.0374$ & $%
-1.0275_{- 0.0318- 0.0509}^{+ 0.0228+ 0.0603}$ & $-1.0134$ \\ 
$\xi$ & $[0,1]$ & $0.0360_{- 0.0360- 0.0360}^{+ 0.0091+ 0.0507}$ & $0.0436$
& $0.0433_{- 0.0433- 0.0433}^{+ 0.0062+ 0.0744}$ & $0.0086$ & $0.1064_{-
0.1064- 0.1064}^{+ 0.0437+ 0.1413}$ & $0.1080$ \\ \hline
$H_0$ & $73.24\pm1.74$ & $68.4646_{- 0.7380- 1.3616}^{+ 0.8199+ 1.3348}$ & $%
68.1714$ & $68.5099_{- 0.9264- 1.7640}^{+ 0.8529+ 2.0520}$ & $68.6939$ & $%
68.5420_{- 0.6763- 1.4114}^{+ 0.7817+ 1.3760}$ & $68.3716$ \\ 
$\Omega_{m0}$ & $-$ & $0.3014_{- 0.0077- 0.0141}^{+ 0.0070+ 0.0139}$ & $%
0.3042$ & $0.3008_{- 0.0078- 0.0163}^{+ 0.0082+ 0.0155}$ & $0.2997$ & $%
0.3030_{- 0.0062- 0.0124}^{+ 0.0063+ 0.0126}$ & $0.3014$ \\ 
$\sigma_8$ & $-$ & $0.8156_{- 0.0137- 0.0244}^{+ 0.0121+ 0.0246}$ & $0.8249$
& $0.8166_{- 0.0166- 0.0280}^{+ 0.0134+ 0.0300}$ & $0.8096$ & $0.8051_{-
0.0185- 0.0396}^{+ 0.0231+ 0.0336}$ & $0.8068$ \\ \hline\hline
\end{tabular}%
\caption{\textit{The table summarizes the mean values of the free and
derived cosmological parameters with their errors at 68.3\% and 95.4\%
confidence regions for IDE 1: $Q=3H\protect\xi(1+w_x)\protect\rho_x$, IDE 2: 
$Q=3H\protect\xi(1+w_x)\protect\rho_c\protect\rho_x/(\protect\rho_c+\protect%
\rho_x)$, and IDE 3: $Q=3H\protect\xi(1+w_x)\protect\rho^2_x/\protect\rho_c$
using the combined analysis CMB $+$ BAO $+$ JLA $+$ RSD $+$ WL $+$ CC $+$ $%
H_0$. We note that, $\Omega_{m0}= \Omega_{c0}+\Omega_{b0}$. } }
\label{tab:results}
\end{table}

\begin{figure}[tbh]
\includegraphics[height=6.7cm, width=7.8cm]{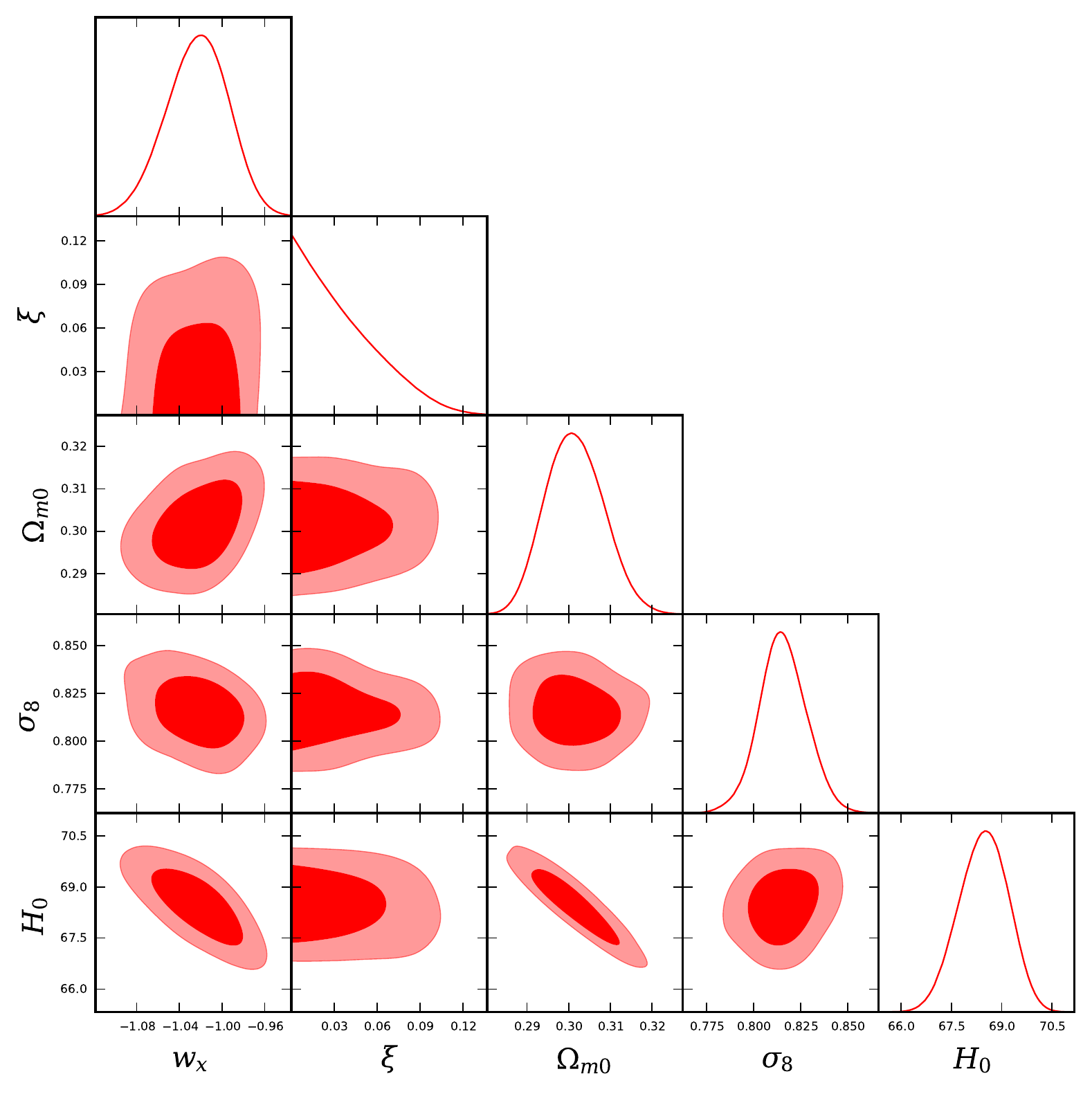}
\caption{\textit{The figure displays the 68.3\% and 95.4\% confidence-region
contour plots for IDE 1 using the combined analysis CMB $+$ BAO $+$ JLA $+$
RSD $+$ WL $+$ CC $+$ $H_0$. Here, $\Omega_{m0}= \Omega_{c0}+ \Omega_{b0}$. }
}
\label{fig:contour1}
\end{figure}

\begin{figure}[tbh]
\includegraphics[width=0.3\textwidth]{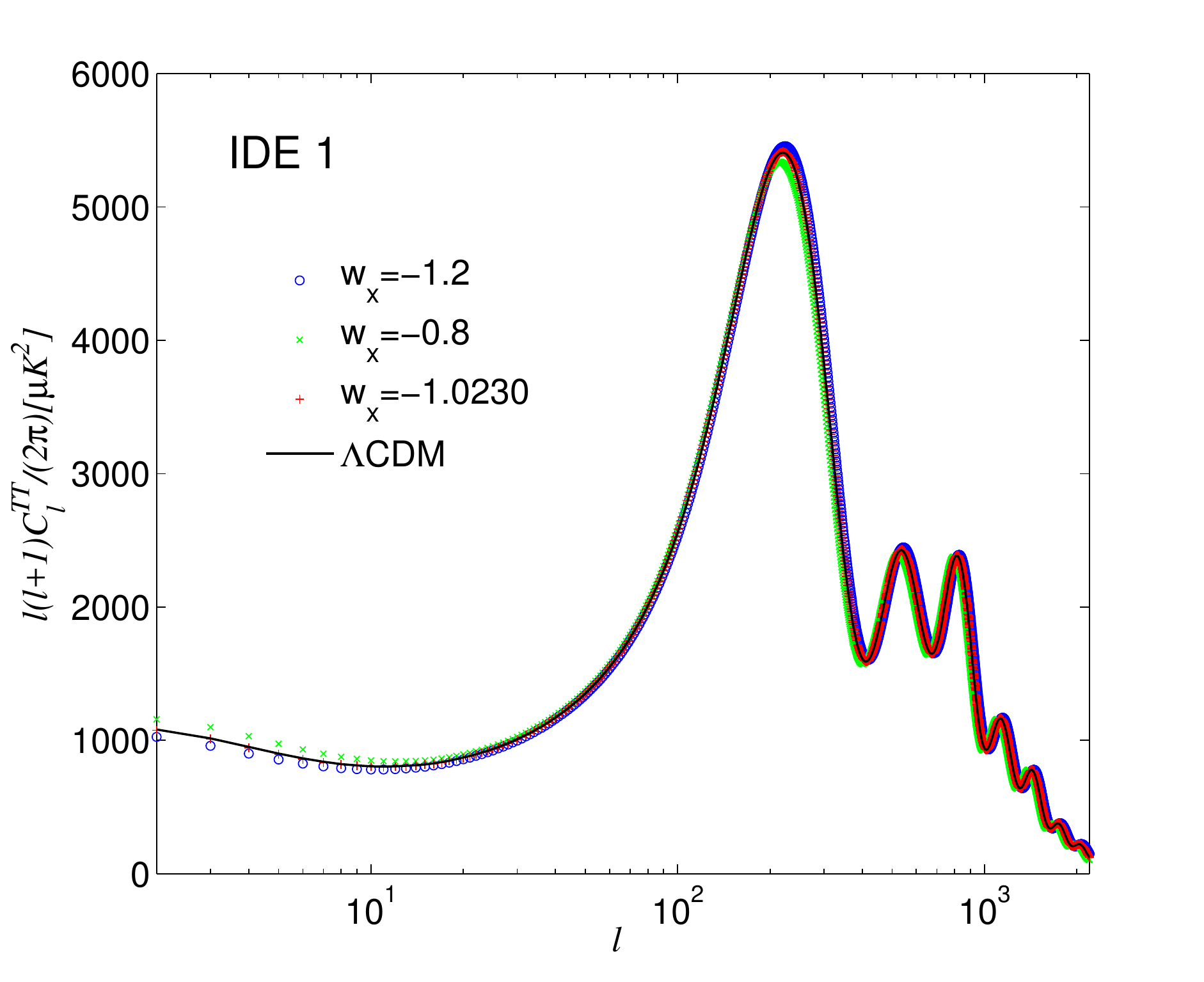} %
\includegraphics[width=0.3\textwidth]{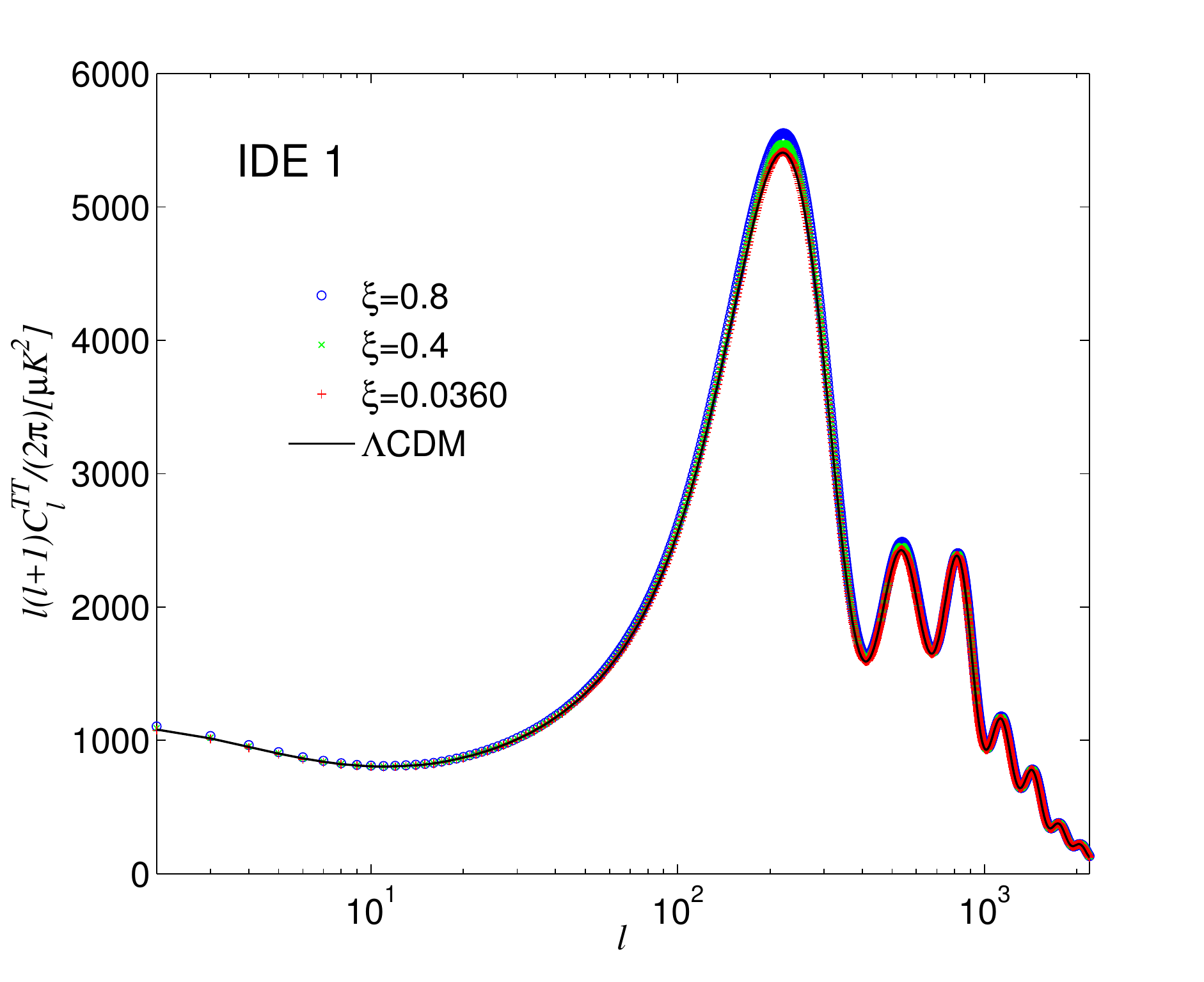}
\caption{\textit{The plots show the angular CMB temperature power spectra of
IDE 1 in compared to the standard $\Lambda$CDM cosmology using the combined
analysis CMB $+$ BAO $+$ JLA $+$ RSD $+$ WL $+$ CC $+$ $H_0$. In the left
panel we show different angular CMB spectra for different values of $w_x$
including its mean value obtained from the above combined analysis while the
right panel shows replica of the left panel but for different values of the
coupling parameter $\protect\xi$ including its mean value from the same
combined analysis.}}
\label{fig:cmbplot1}
\end{figure}
\end{center}

\begin{figure}[tbh]
\includegraphics[width=0.3\textwidth]{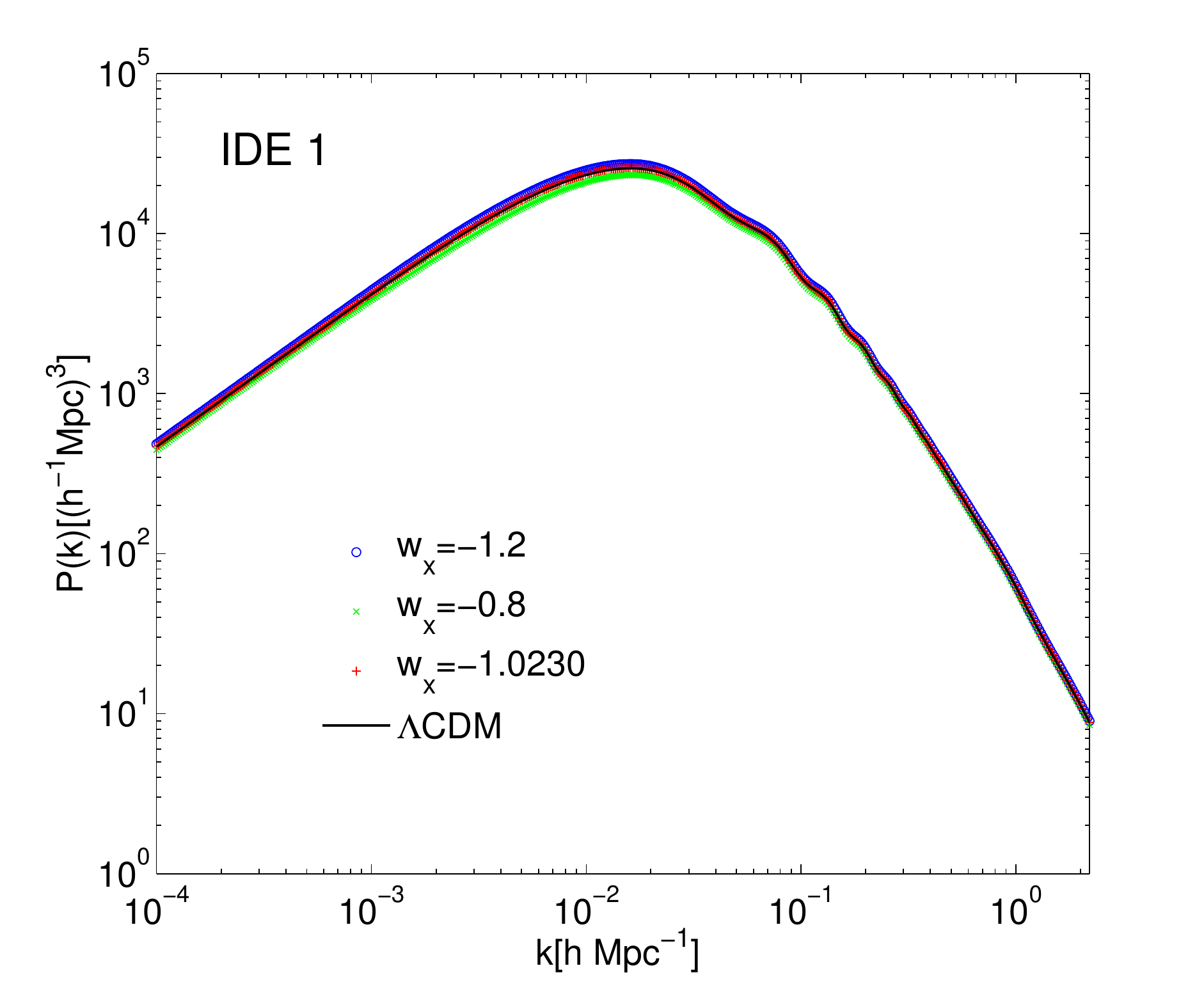} %
\includegraphics[width=0.3\textwidth]{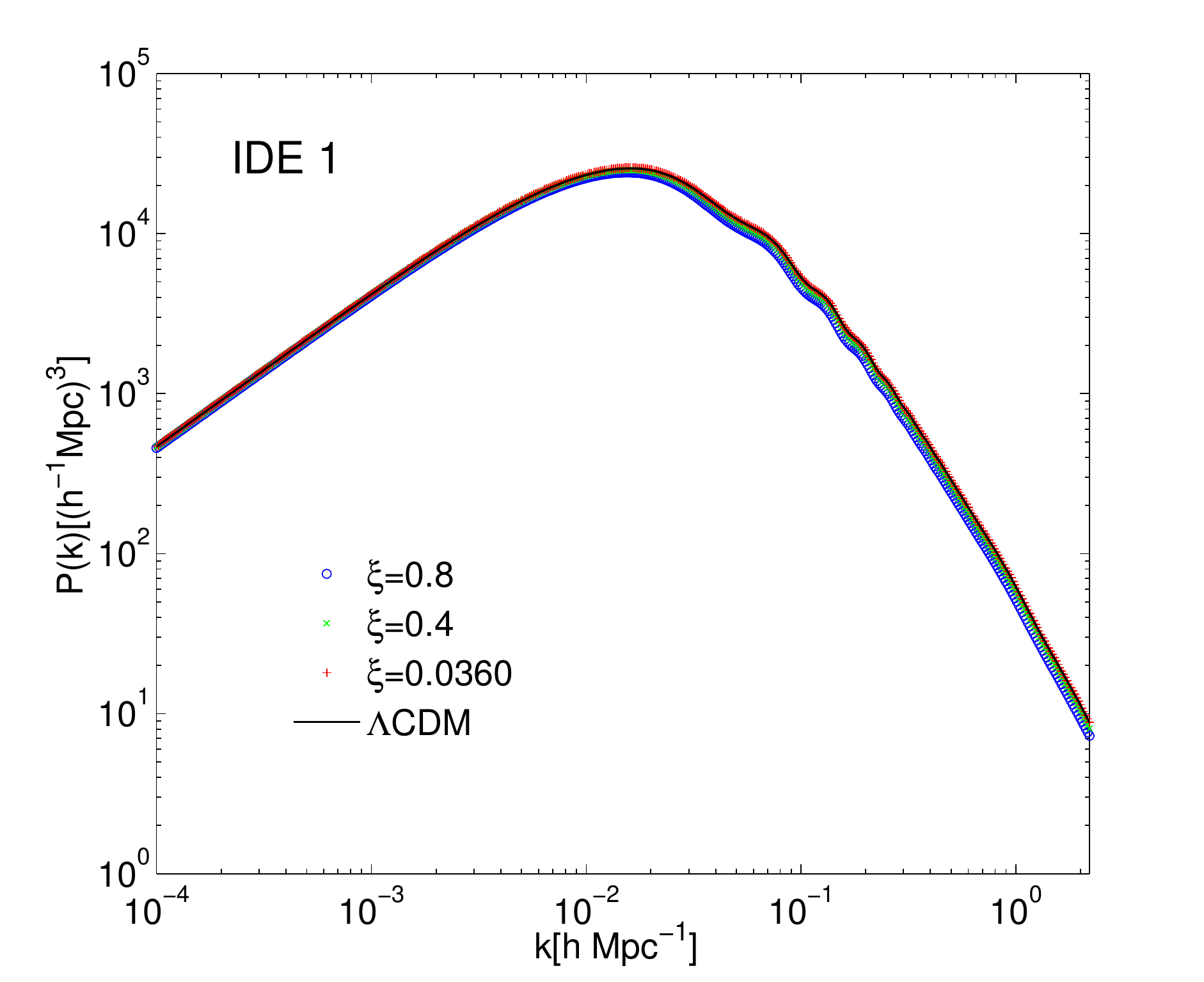}
\caption{\textit{The figure shows the behavior of the matter power spectra
of IDE 1 in compared to the $\Lambda$CDM cosmology for the combined
observational analysis CMB $+$ BAO $+$ JLA $+$ RSD $+$ WL $+$ CC $+$ $H_0$.
In the left panel we use different values of the dark energy equation of
state $w_x$, while in the right panel we vary the coupling parameter $%
\protect\xi$. }}
\label{fig:Mpower1}
\end{figure}
\begin{figure}[tbh]
\includegraphics[width=0.34\textwidth]{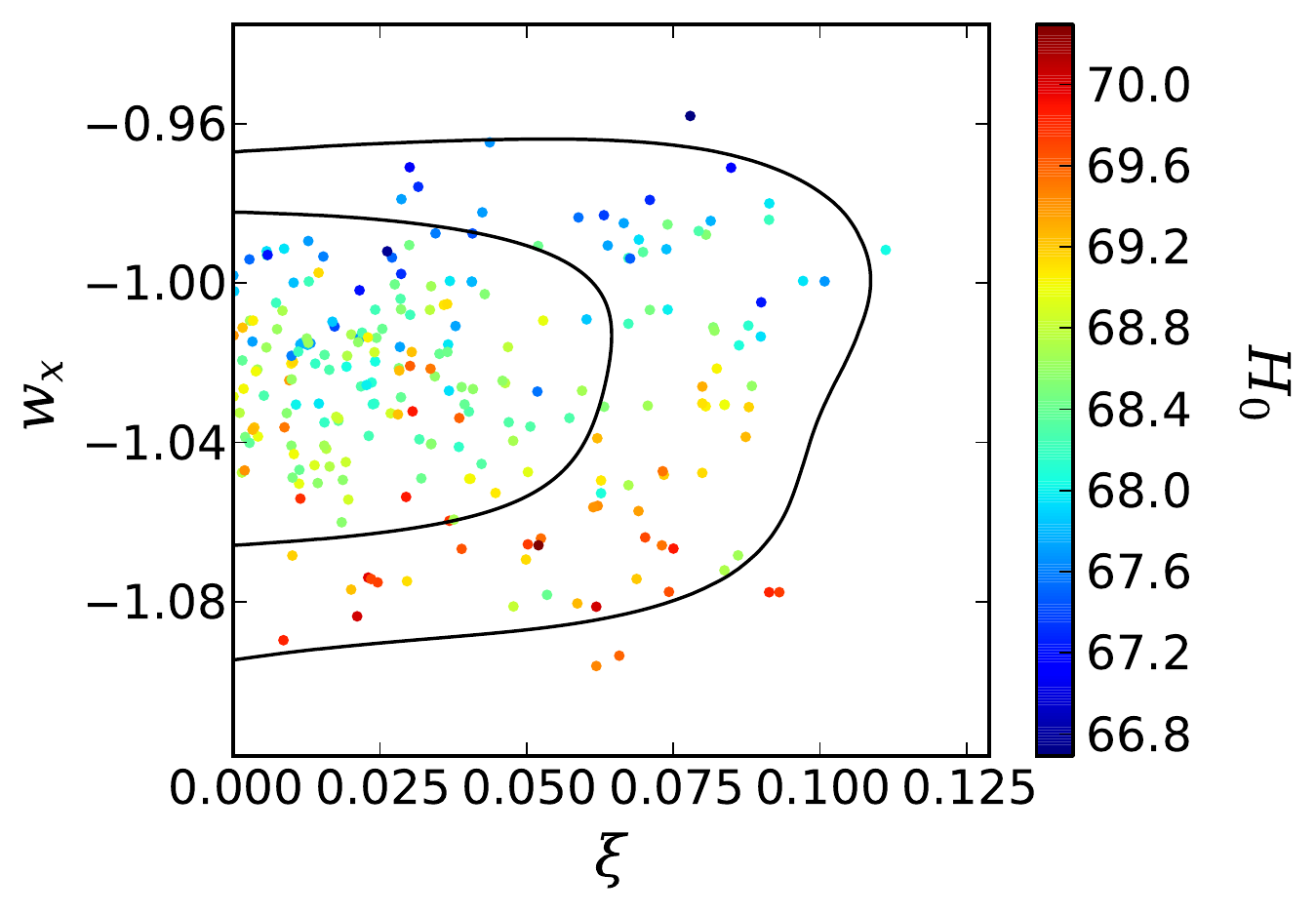}
\caption{\textit{MCMC samples in the $(w_x, \protect\xi)$ plane coloured by
the Hubble constant value $H_0$ for IDE 1 analyzed with the combined
analysis CMB $+$ BAO $+$ JLA $+$ RSD $+$ WL $+$ CC $+$ $H_0$. } }
\label{fig:scatterIDE1}
\end{figure}

\begin{center}
\begin{figure}[tbh]
\includegraphics[height=6.7cm, width=7.8cm]{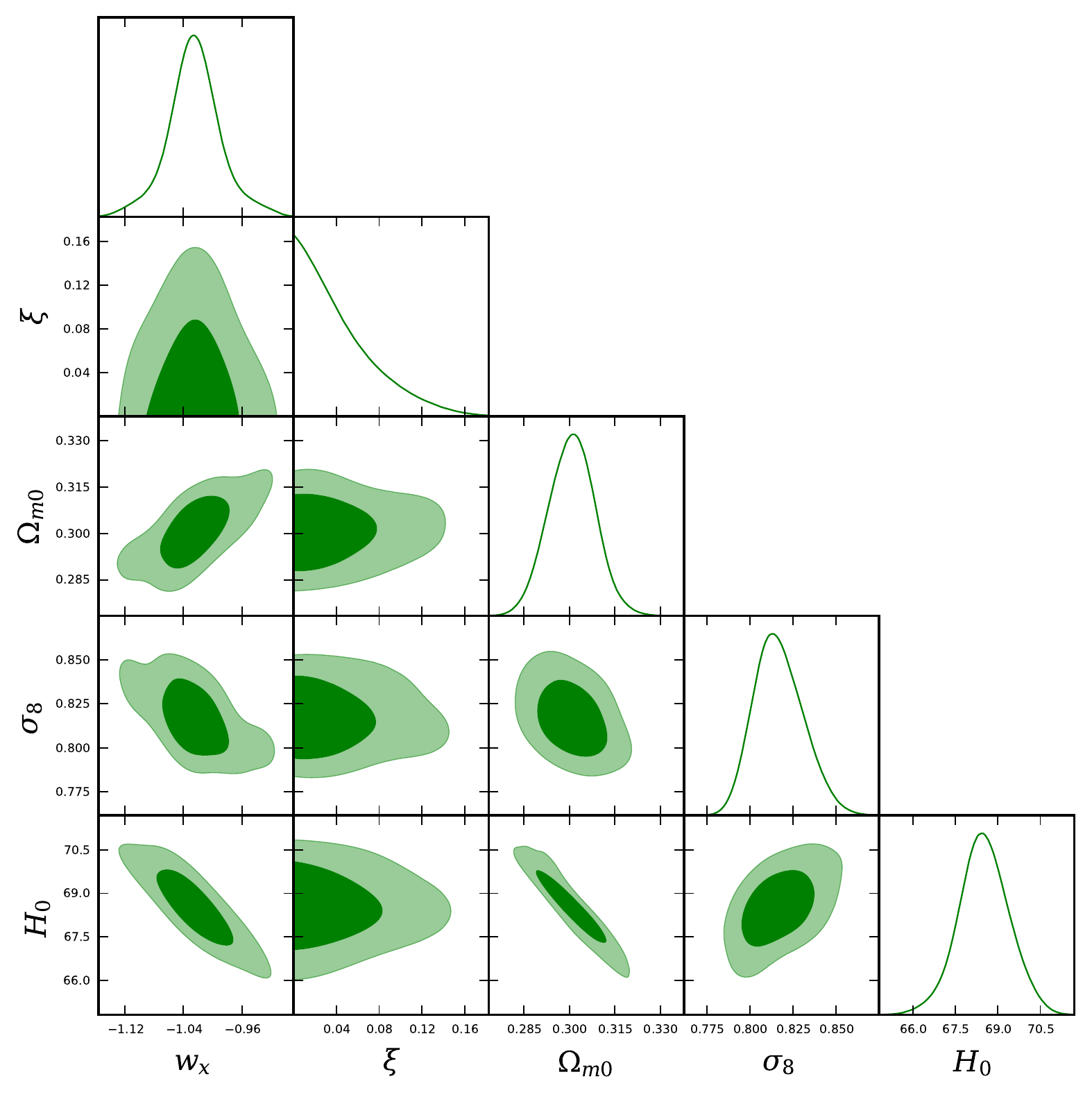}
\caption{\textit{The figure displays the 68.3\% and 95.4\% confidence-region
contour plots for different combinations of the free parameters of IDE 2
using the combined analysis CMB $+$ BAO $+$ JLA $+$ RSD $+$ WL $+$ CC $+$ $%
H_0$. Here, $\Omega_{m0}= \Omega_{c0}+ \Omega_{b0}$.} }
\label{fig:contour2}
\end{figure}

\begin{figure}[tbh]
\includegraphics[width=0.3\textwidth]{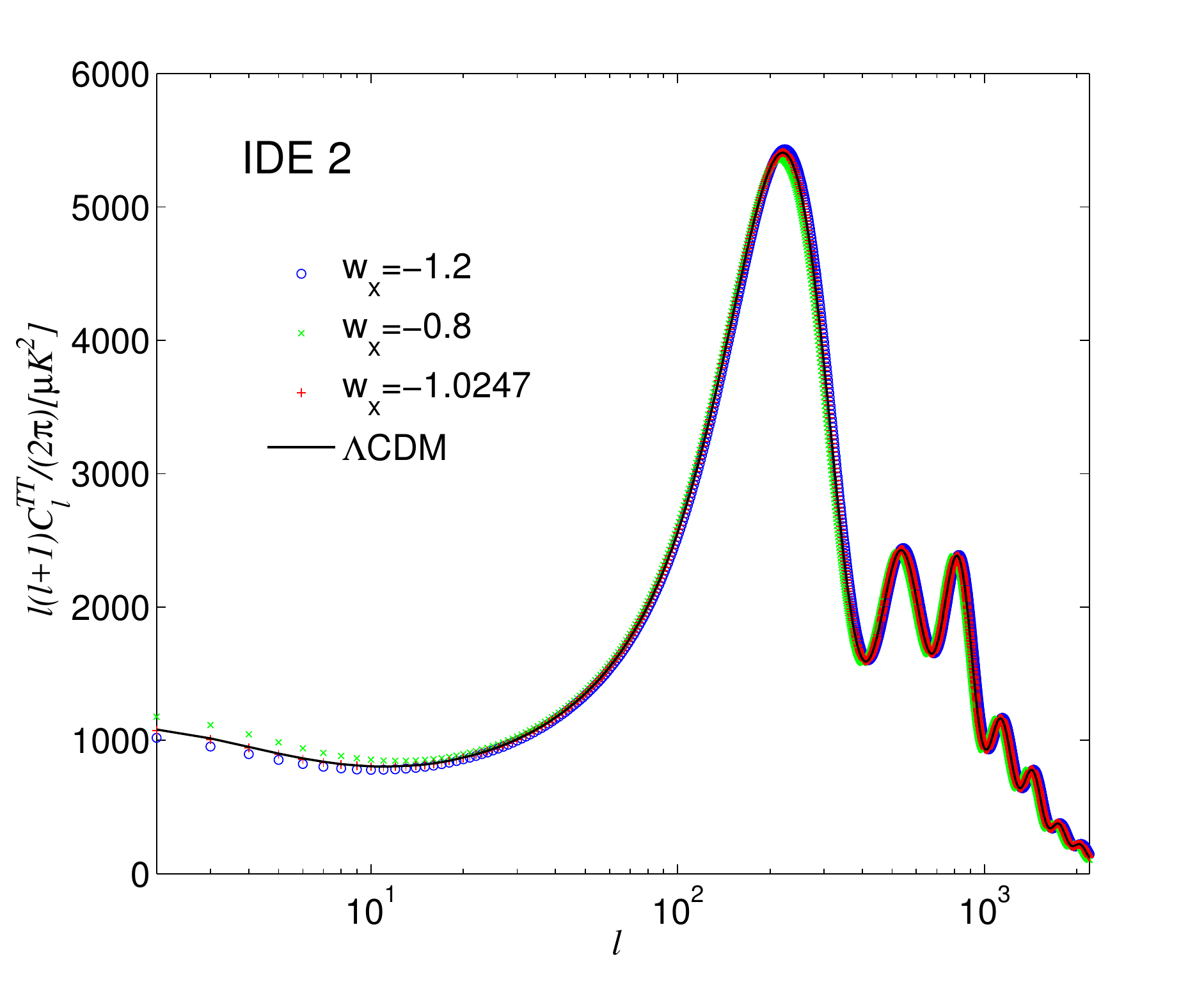} %
\includegraphics[width=0.3\textwidth]{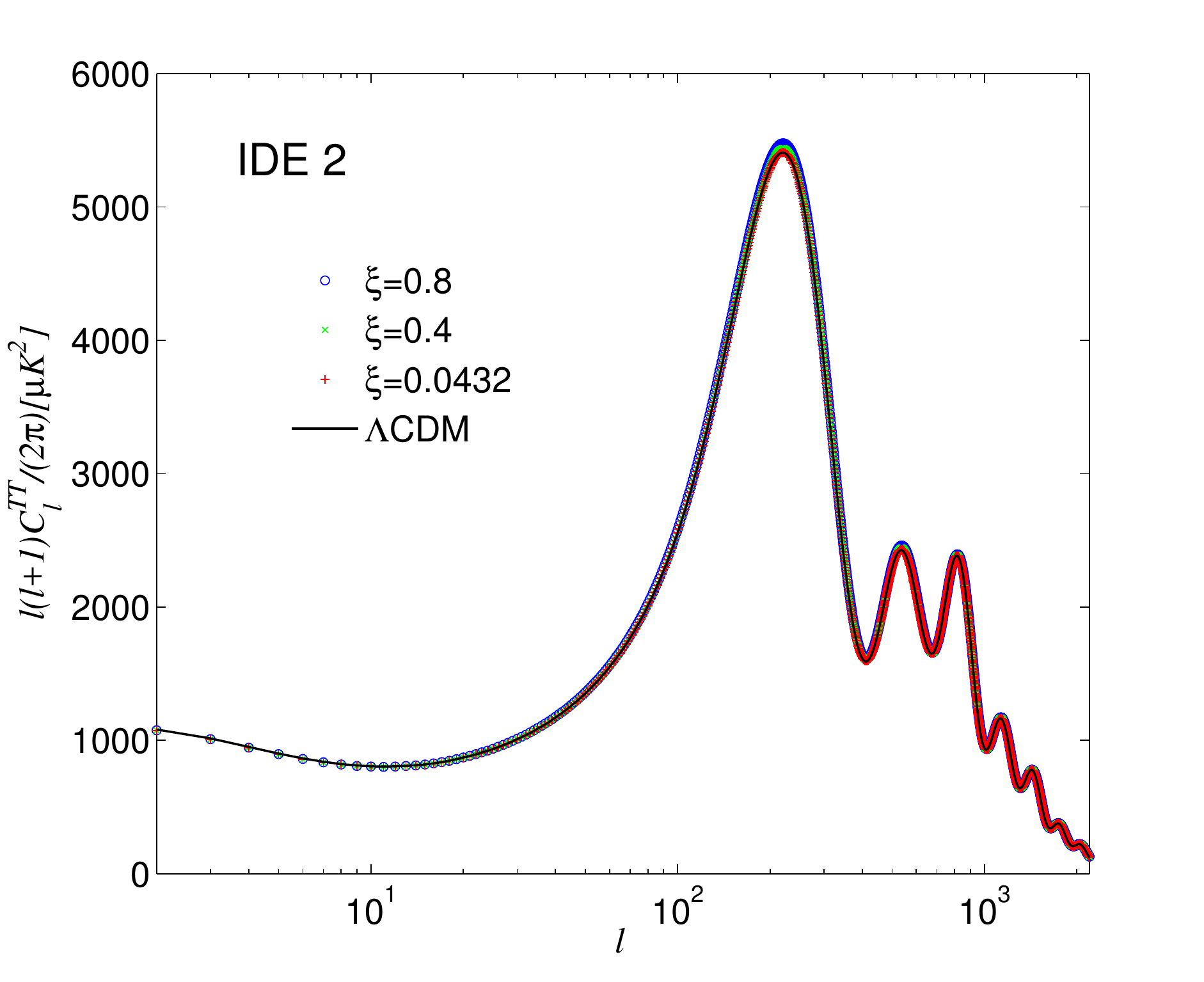}
\caption{\textit{The plots show the angular CMB temperature power spectra of
IDE 2 in compared to the standard $\Lambda$CDM cosmology using the combined
analysis CMB $+$ BAO $+$ JLA $+$ RSD $+$ WL $+$ CC $+$ $H_0$. In the left
panel we show different angular CMB spectra for different values of $w_x$
including its mean value obtained from the combined analysis while the right
panel shows replica of the left panel but for different values of the
coupling parameter $\protect\xi$ including its mean value from the same
combined analysis.}}
\label{fig:cmbplot2}
\end{figure}

\begin{figure}[tbh]
\includegraphics[width=0.3\textwidth]{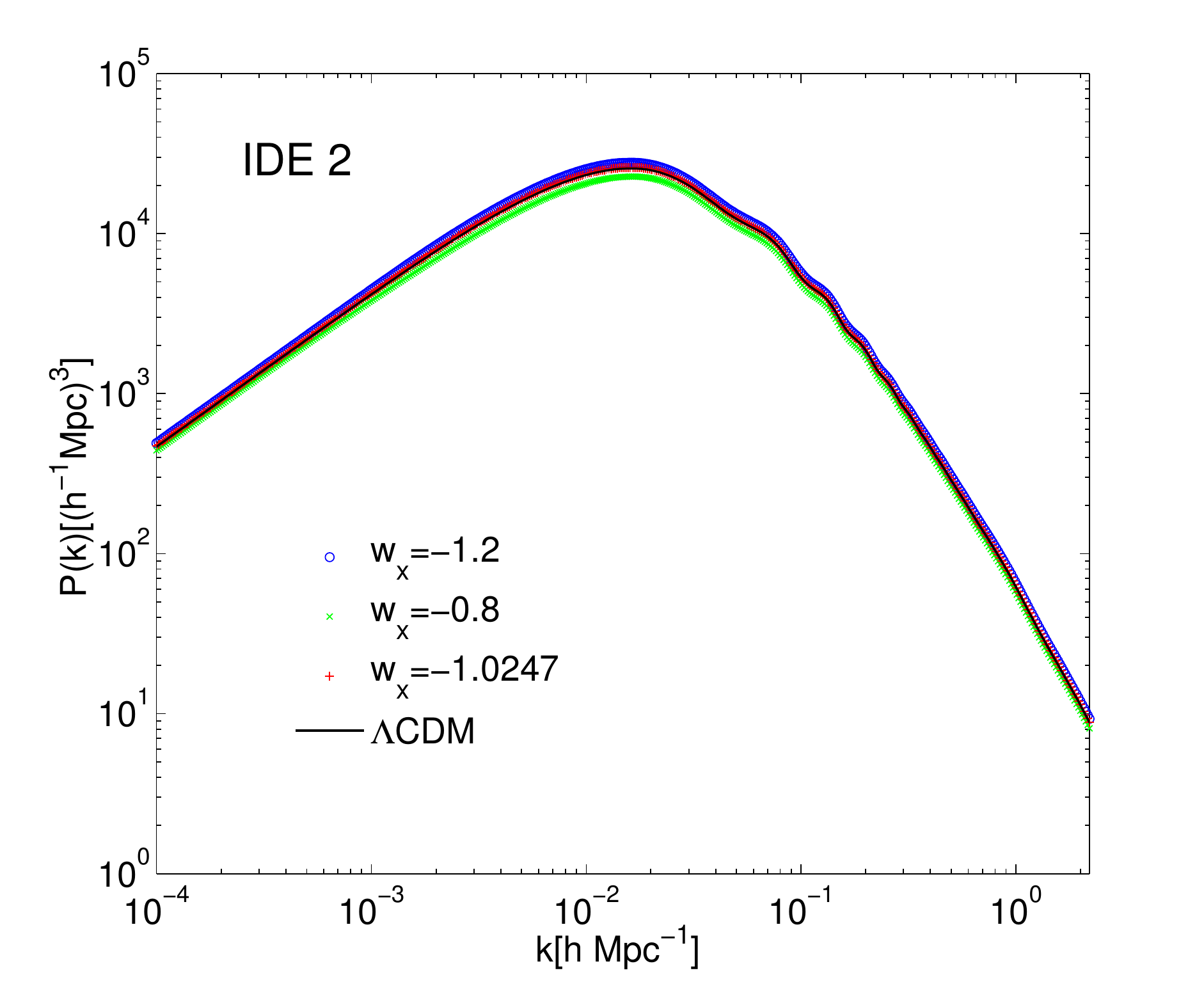} %
\includegraphics[width=0.3\textwidth]{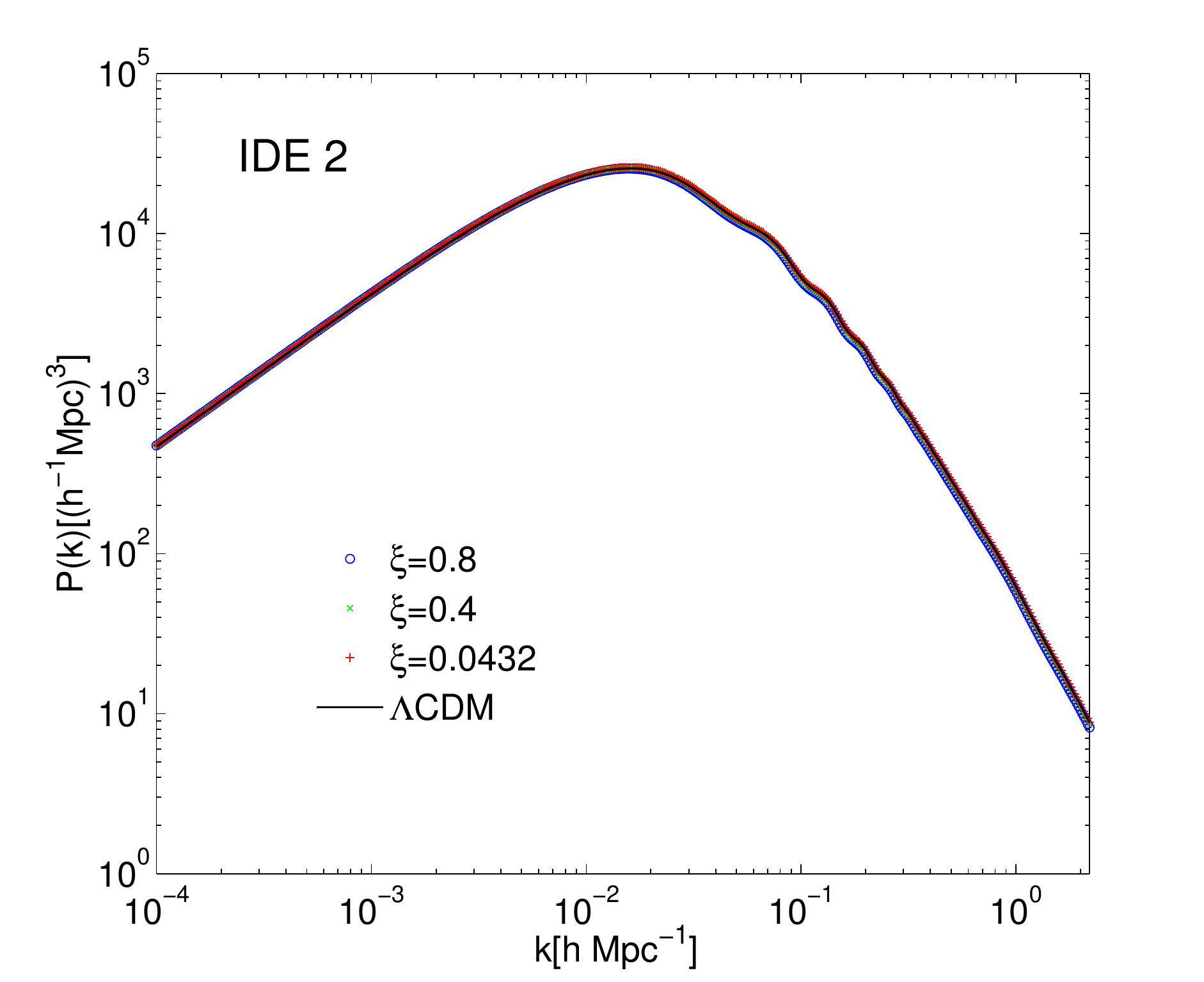}
\caption{\textit{The figure shows the behavior of the matter power spectra
of IDE 2 in compared to the $\Lambda$CDM cosmology for CMB $+$ BAO $+$ JLA $%
+ $ RSD $+$ WL $+$ CC $+$ $H_0$. In the left panel we use different values
of the dark energy equation of state $w_x$, while in the right panel we vary
the coupling parameter $\protect\xi$.}}
\label{fig:Mpower2}
\end{figure}

\begin{figure}[tbh]
\includegraphics[width=0.34\textwidth]{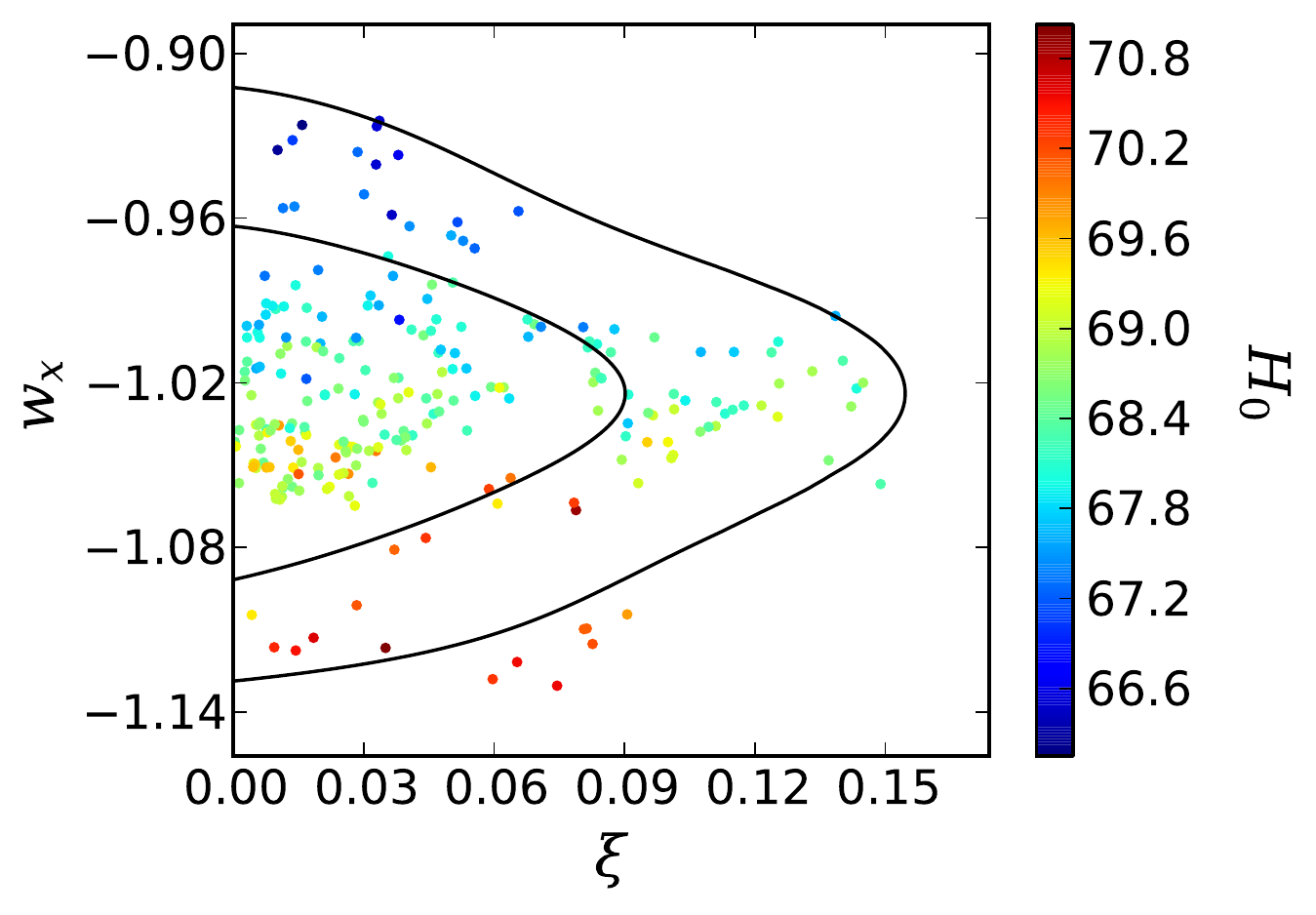}
\caption{\textit{MCMC samples in the $(w_x, \protect\xi)$ plane coloured by
the Hubble constant value $H_0$ for IDE 2 analyzed with the combined
analysis CMB $+$ BAO $+$ JLA $+$ RSD $+$ WL $+$ CC $+$ $H_0$.} }
\label{fig:scatterIDE2}
\end{figure}

\begin{figure}[tbh]
\includegraphics[height=6.7cm, width=7.8cm]{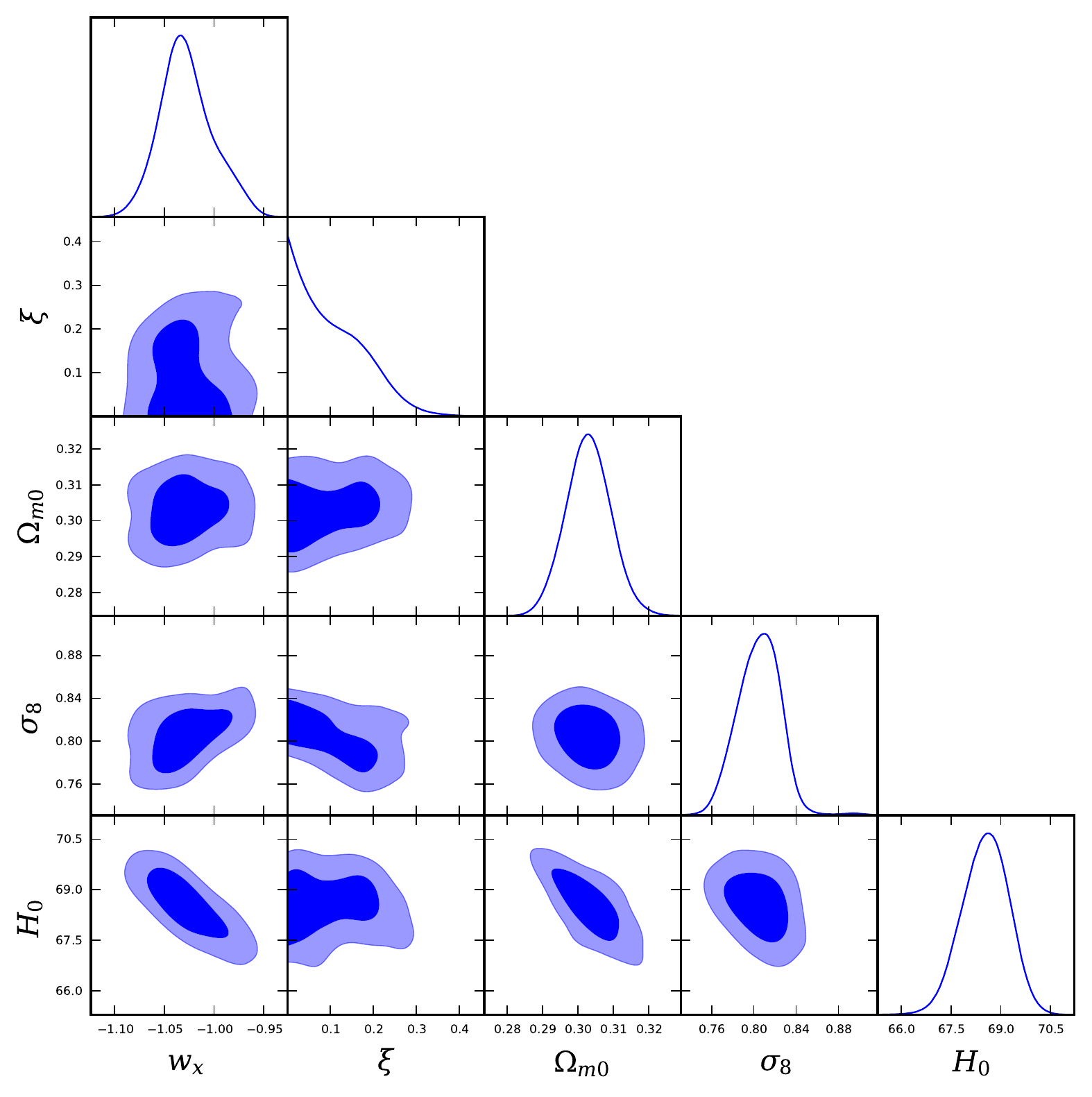}
\caption{\textit{The figure displays the 68.3\% and 95.4\% confidence-region
contour plots for IDE 3 using the combined analysis CMB $+$ BAO $+$ JLA $+$
RSD $+$ WL $+$ CC $+$ $H_0$. Here, $\Omega_{m0}= \Omega_{c0}+ \Omega_{b0}$. }
}
\label{fig:contour3}
\end{figure}

\begin{figure}[tbh]
\includegraphics[width=0.3\textwidth]{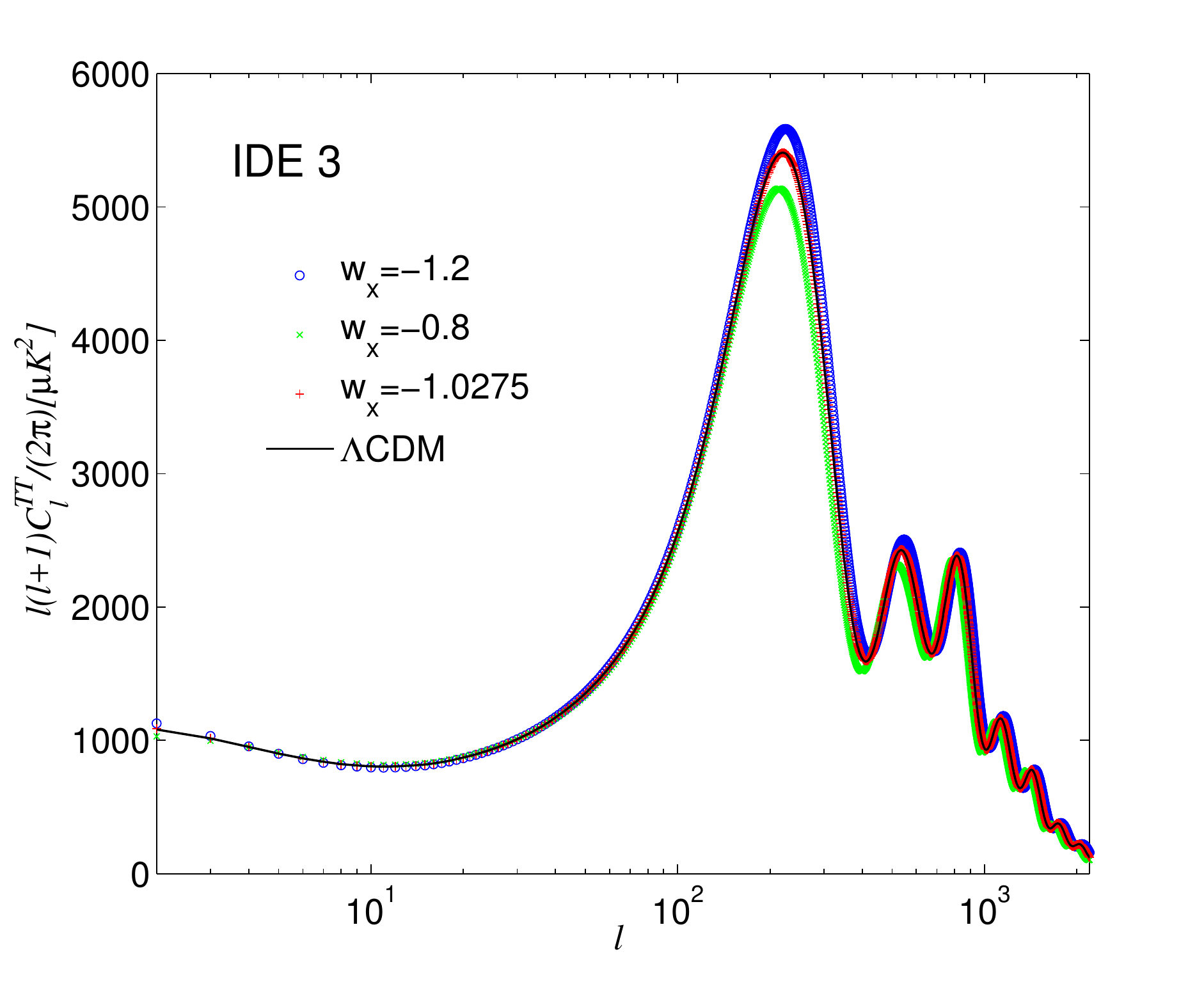} %
\includegraphics[width=0.3\textwidth]{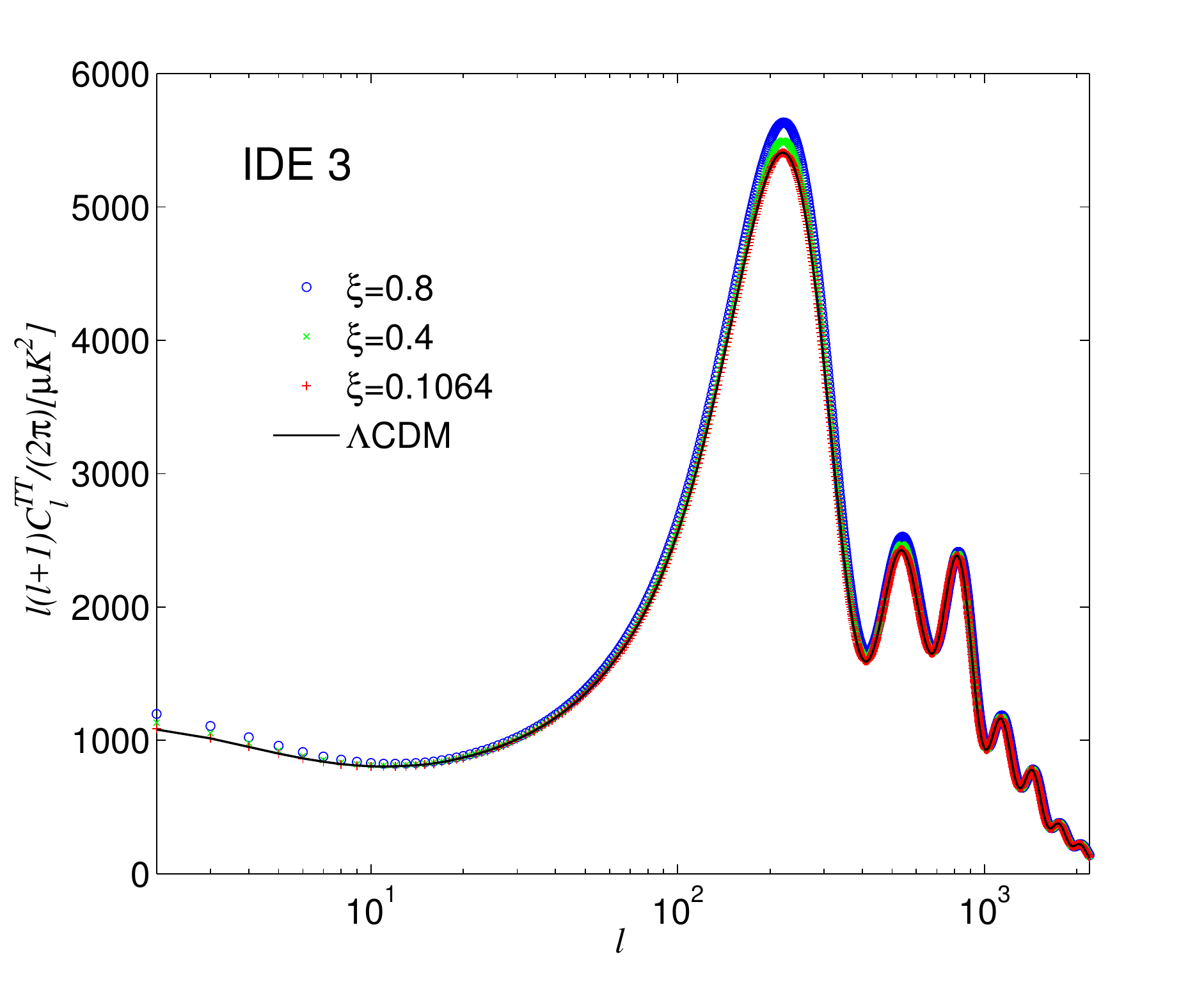}
\caption{\textit{The plots show the angular CMB temperature power spectra of
IDE 3 in compared to the standard $\Lambda$CDM cosmology for the analysis
CMB $+$ BAO $+$ JLA $+$ RSD $+$ WL $+$ CC $+$ $H_0$. In the left panel we
show different angular CMB spectra for different values of $w_x$ including
its mean value obtained from the combined analysis while the right panel
shows replica of the left panel but for different values of the coupling
parameter $\protect\xi$ including its mean value from the same combined
analysis.} }
\label{fig:cmbplot3}
\end{figure}

\begin{figure}[tbh]
\includegraphics[width=0.3\textwidth]{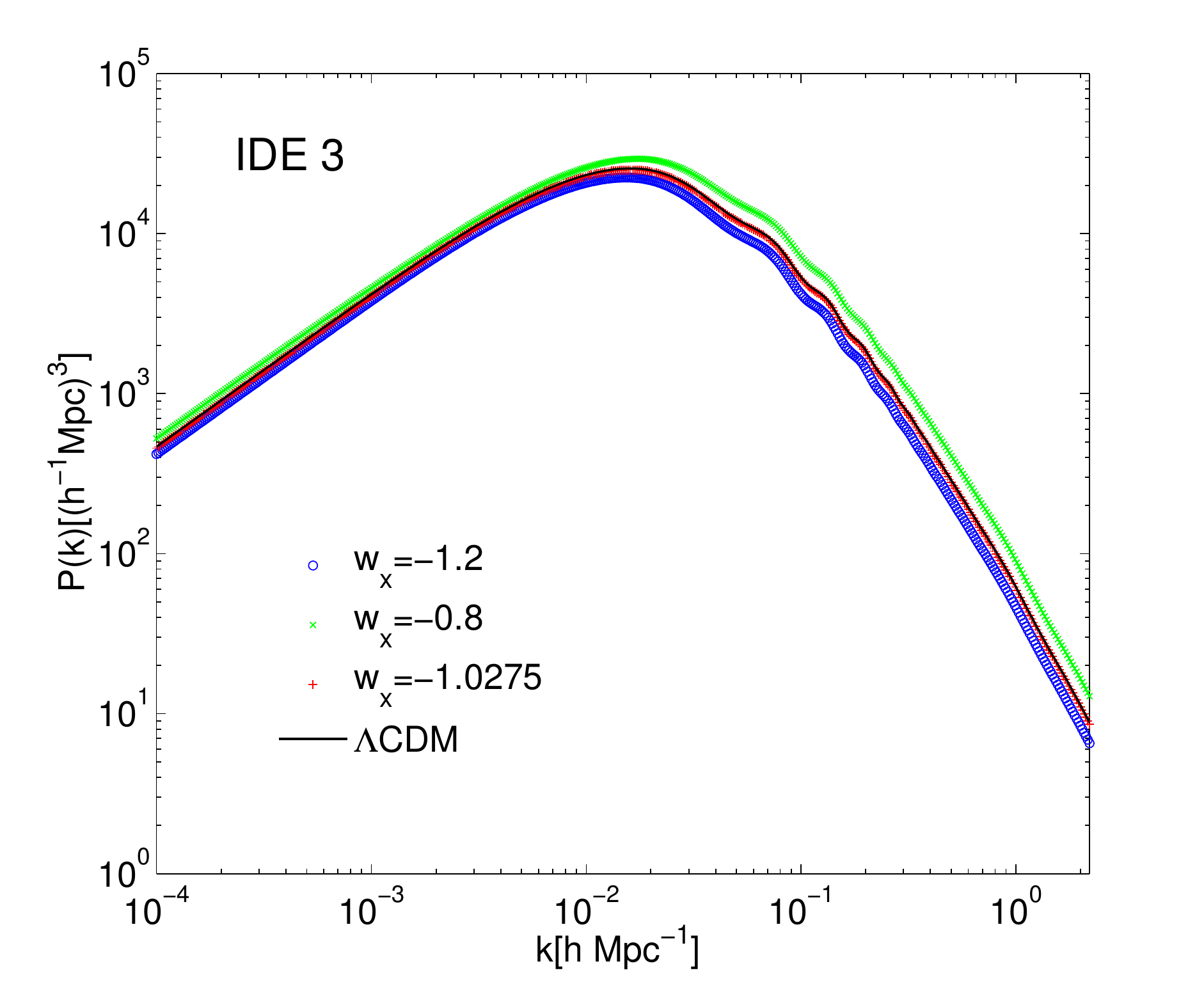} %
\includegraphics[width=0.3\textwidth]{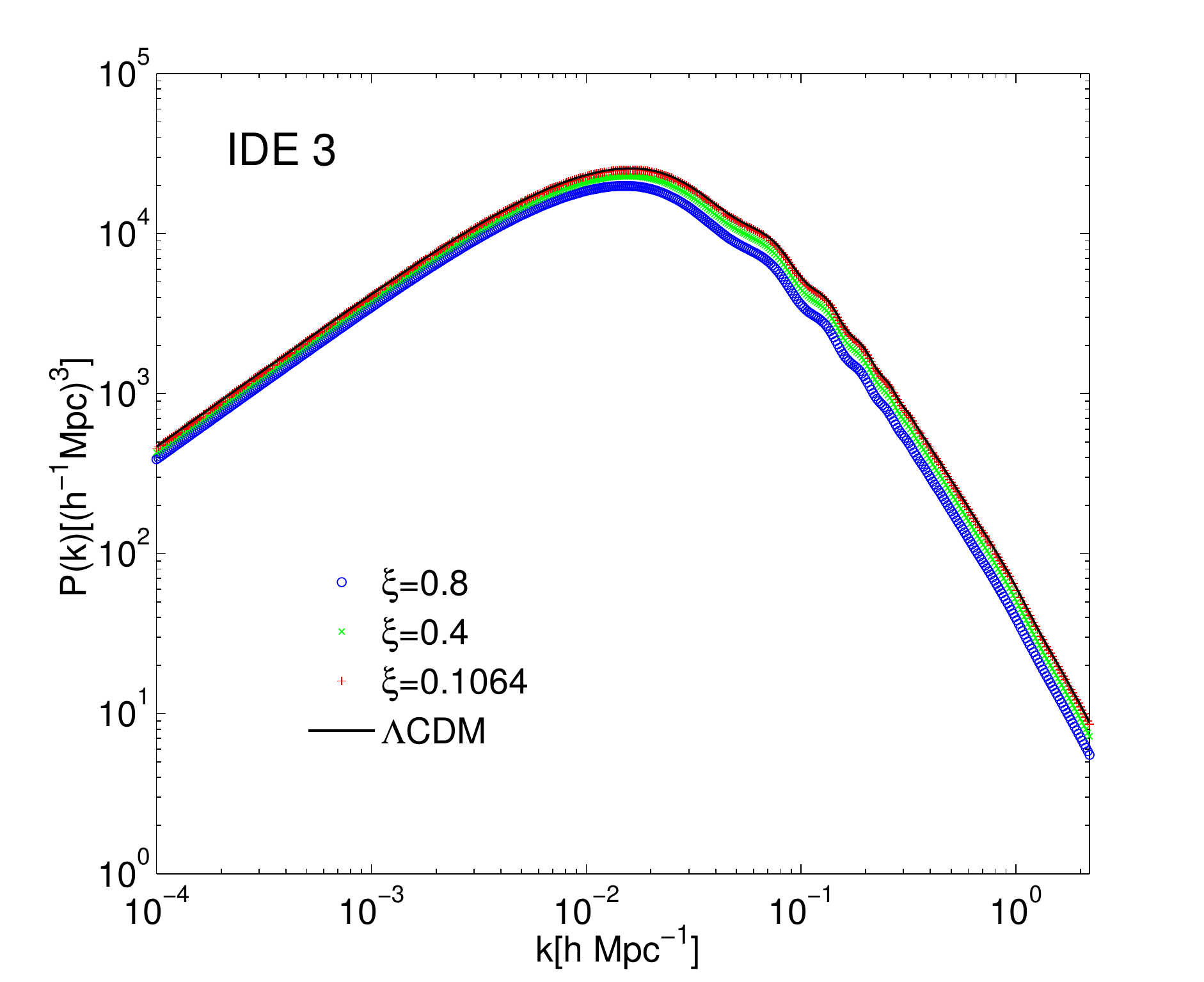}
\caption{\textit{The figure shows the behavior of the matter power spectra
of IDE 3 in compared to the $\Lambda$CDM cosmology for CMB $+$ BAO $+$ JLA $%
+ $ RSD $+$ WL $+$ CC $+$ $H_0$. In the left panel we use different values
of the dark energy equation of state $w_x$, while in the right panel we vary
the coupling parameter $\protect\xi$.}}
\label{fig:Mpower3}
\end{figure}

\begin{figure}[tbh]
\includegraphics[width=0.34\textwidth]{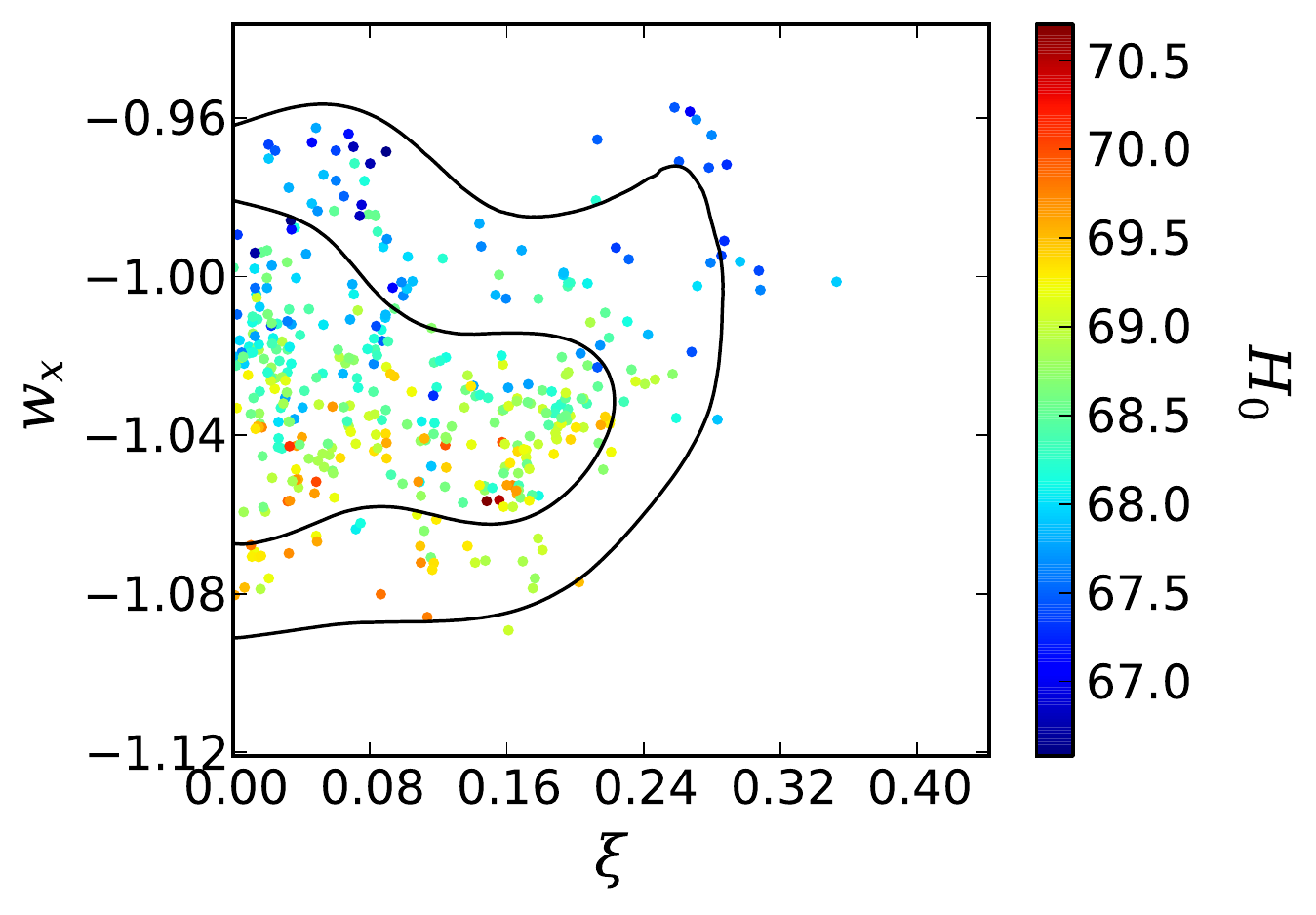}
\caption{\textit{MCMC samples in the $(w_x, \protect\xi)$ plane coloured by
the Hubble constant value $H_0$ for IDE 3 analyzed with the combined
analysis CMB $+$ BAO $+$ JLA $+$ RSD $+$ WL $+$ CC $+$ $H_0$.} }
\label{fig:scatterIDE3}
\end{figure}
\end{center}

\begin{itemize}
\item \textbf{IDE 1}: In Fig. \ref{fig:contour1}, we display the 68.3\% and
95.4\% confidence-level (C.L.) contour plots for different
combinations of the free parameters of this model 
as well as the one-dimensional posterior distribution for each parameter
From the analysis, one finds that the model predicts a very small coupling in the dark
sectors, with $\xi =0.0360_{-0.0360}^{+0.0091}$ at 68.3\%
confidence-level (CL). Also, as one can see, a zero value for $\xi 
$ (i.e. no interaction) is allowed at 68.3\% CL. This implies that
within 68.3\% CL, our interaction model is can recover the non-interacting $%
w_{x}$CDM model.  But, our analysis also shows that the equation
of state of dark energy, $w_{x}$, can cross the phantom dividing %
line, with $w_{x}=-1.0230_{-0.0257}^{+0.0329}$ at 68.3\% CL with the best
fit value $w_{x}=-1.0210$. Although, at the 68.3\% CL, $w_{x}$ could be
greater than `$-1$', this means that its quintessential character cannot be
excluded -- at least according to the current observational data
employed in this analysis. However, we note that numerical values
of both mean and best fit values of $w_{x}$, are close to the cosmological
constant limit of $ w_x=-1$. Thus, from the
constraints on the coupling strength, as well as the equation of state for
dark energy, one finds that the observational data favor a very small
interaction in the dark sector and the model for the background evolution
displays a close match to the $\Lambda $CDM cosmology. We also find
that, at the perturbative level, IDE 1 cannot be distinguished from the $%
\Lambda $CDM cosmology. In Figures \ref{fig:cmbplot1} and \ref{fig:Mpower1},
we have described the angular power spectra of the CMB temperature
anisotropy, and the matter power spectra for different values of $w_{x}$ and 
$\xi $. We see that a very slight deviation is observed at the highest peak
of the plot (right-hand panel of Fig. \ref{fig:cmbplot1}) for a higher
coupling strength of $\xi =0.8$. A very similar observation can be made
about the matter power spectra (right-hand panel of \ref{fig:Mpower1}) for $%
\xi =0.8$. However, overall, the model does not show any significant
deviation from $\Lambda $CDM even for such a high coupling strength.
Similarly, as $w_{x}$ deviates from `$-1$' towards the quintessence regime,
a very slight deviation from the $\Lambda $CDM cosmology is observed,
although it is not significant either. Further, in Figure \ref%
{fig:scatterIDE1}, we have displayed the two-dimensional marginalized
posterior distribution for the parameters $(w_{x},\xi )$ using the combined
analysis mentioned above. The points in Figure \ref{fig:scatterIDE1} are the
samples from the chains of the combined analysis that have been colored by
the values of $H_{0}$. From this figure, it is seen that the higher values
of $H_{0}$ favor the phantom regime, $w_{x}<-1,$ while the lower values of $%
H_{0}$ favor the quintessence dark energy, i.e. $w_{x}>-1$. In fact, a
shifting from phantom to quintessence dark energy is displayed as the Hubble
parameter values decrease from higher values. Furthermore, we can
also see that a non-zero interaction might be useful to ease the tension on $%
H_{0}$ created by the $\Lambda $CDM-based Planck
estimation ($H_{0}=67.27\pm 0.66$ km~s$^{-1}$~Mpc$^{-1}$) \cite%
{ref:Planck2015-3} and the local measurements by Riess et al. ($%
H_{0}=73.24\pm 1.74$ km~s$^{-1}$~Mpc$^{-1}$) \cite{Riess:2016jrr}. From our
analysis, we find that the introduction of a coupling into the dark
sector gives $%
H_{0}=68.4646_{-0.7380-1.3616-1.8747}^{+0.8199+1.3348+1.6568}$, which shows
that the coupling does produce a shift of the Hubble parameter 
towards higher values, and consequently the tension on $H_{0}$
might be eased in the presence of the interaction. %
The easing of the $H_{0}$ tension in the presence of an
interaction in the dark sector has also been noticed in some earlier works 
\cite{Kumar:2016zpg, Kumar:2017dnp,DiValentino:2017iww} with some different
interactions, and thus it might be considered to be an interesting outcome
of such $w_{x}$CDM$+\xi $ scenario.  One can also see that the $\sigma _{8}$
value extracted from this model matches with the Planck estimation \cite%
{ref:Planck2015-3} when lensing is added to either Planck TT+lowP, or Planck
TT, TE, EE+lowP. This means that the estimated values of $\sigma _{8}$ are, $%
\sigma _{8}=0.8149\pm 0.0093$ (Planck TT+lowP+lensing) \cite%
{ref:Planck2015-3} and $\sigma _{8}=0.8150\pm 0.0087$ (Planck TT, TE,
EE+lowP+lensing) \cite{ref:Planck2015-3}. The external data BAO+JLA+$H_{0}$
added to both these data (Planck TT+lowP and Planck TT, TE, EE+lowP) agree
with the same estimation. \newline
\end{itemize}

\begin{center}
\begin{figure}[tbh]
\includegraphics[height=6.7cm, width=7.8cm]{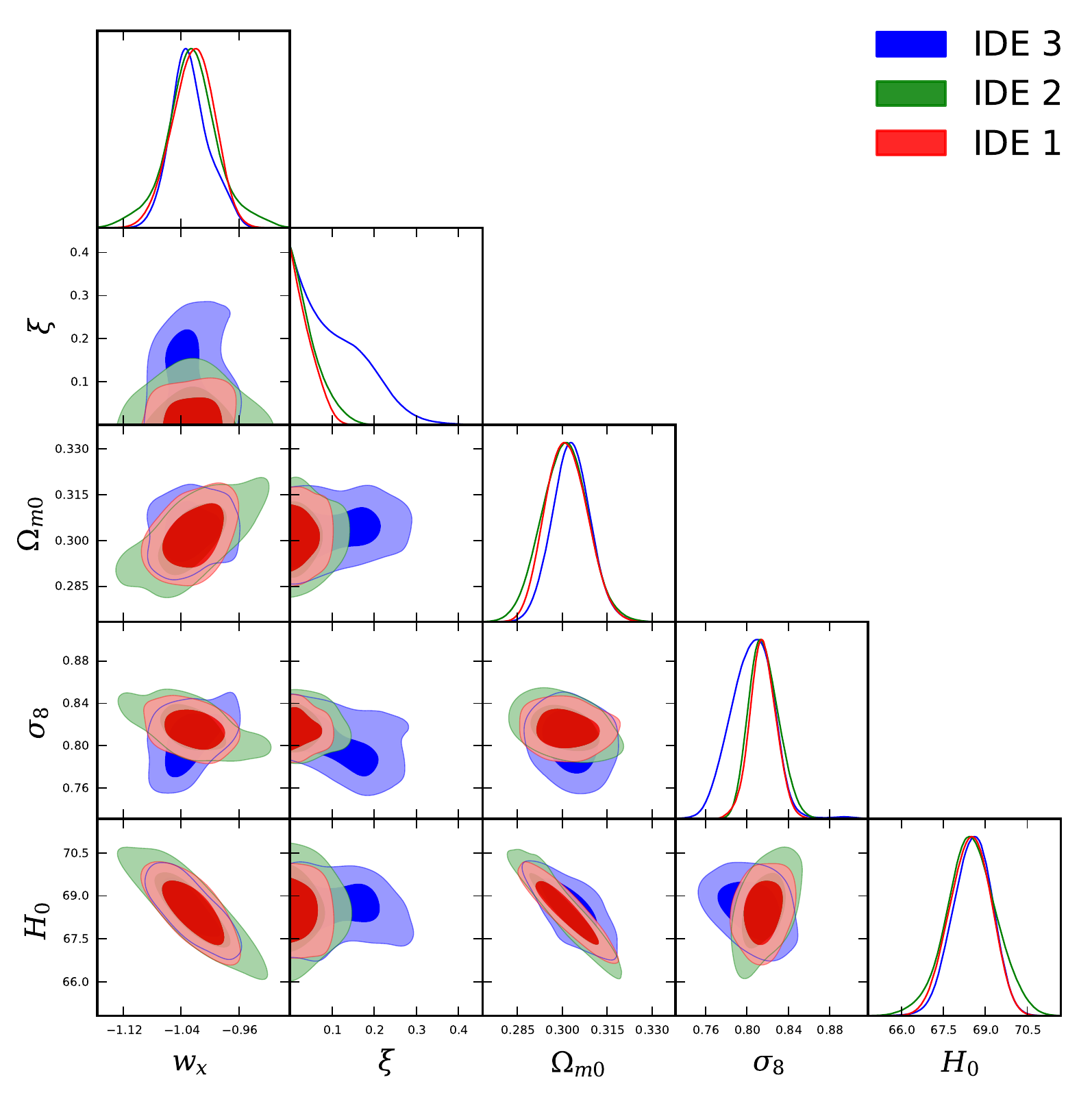}
\caption{\textit{The figure displays the 68.3\% and 95.4\% confidence-region
contour plots for three interacting dark energy models, namely IDE 1, IDE 2
and IDE 3 using the combined analysis CMB $+$ BAO $+$ JLA $+$ RSD $+$ WL $+$
CC $+$ $H_0$. Here $\Omega_{m0}= \Omega_{c0}+ \Omega_{b0}$.} }
\label{fig:contour-all}
\end{figure}

\begin{figure}[tbh]
\includegraphics[width=0.3\textwidth]{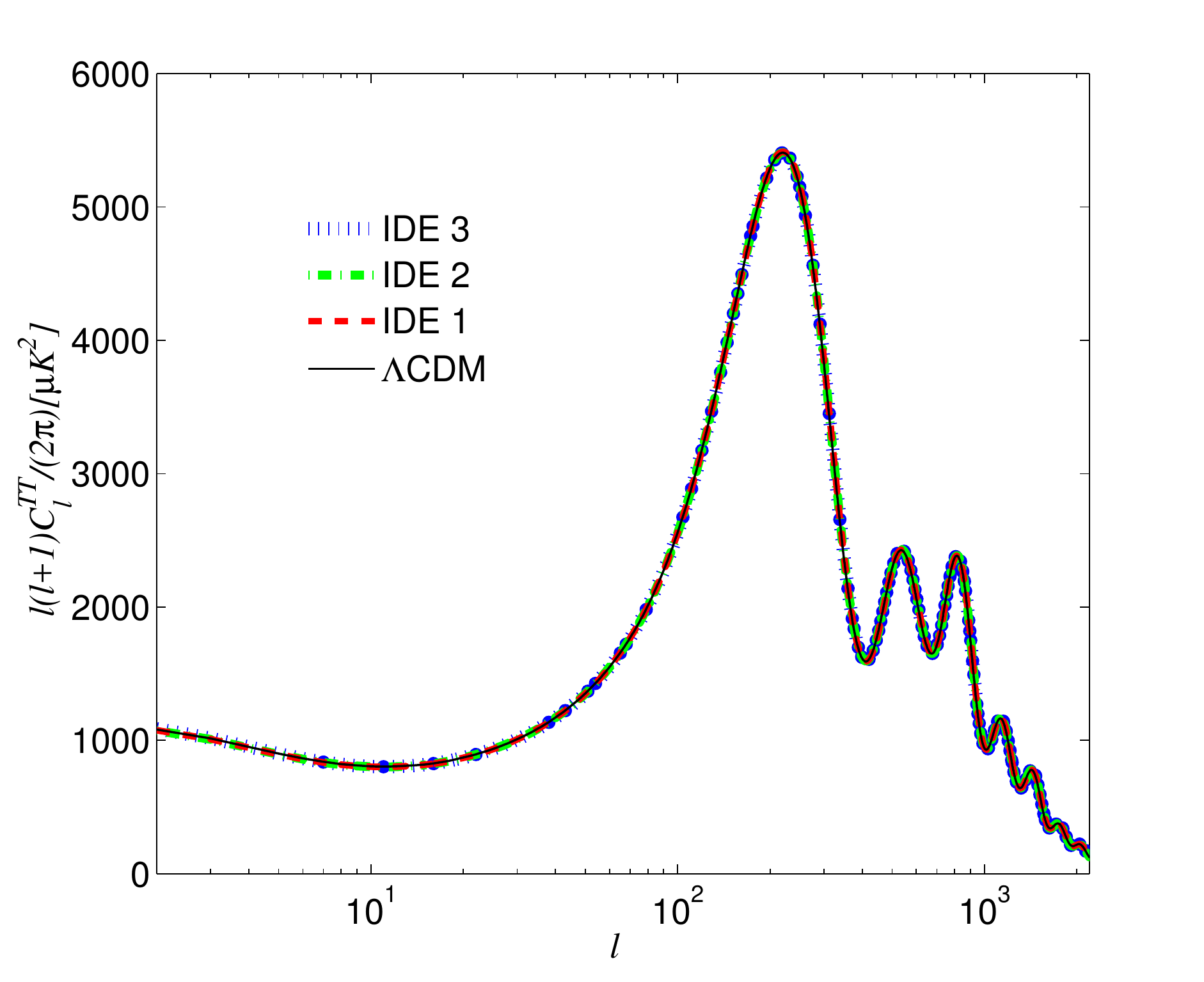} %
\includegraphics[width=0.3\textwidth]{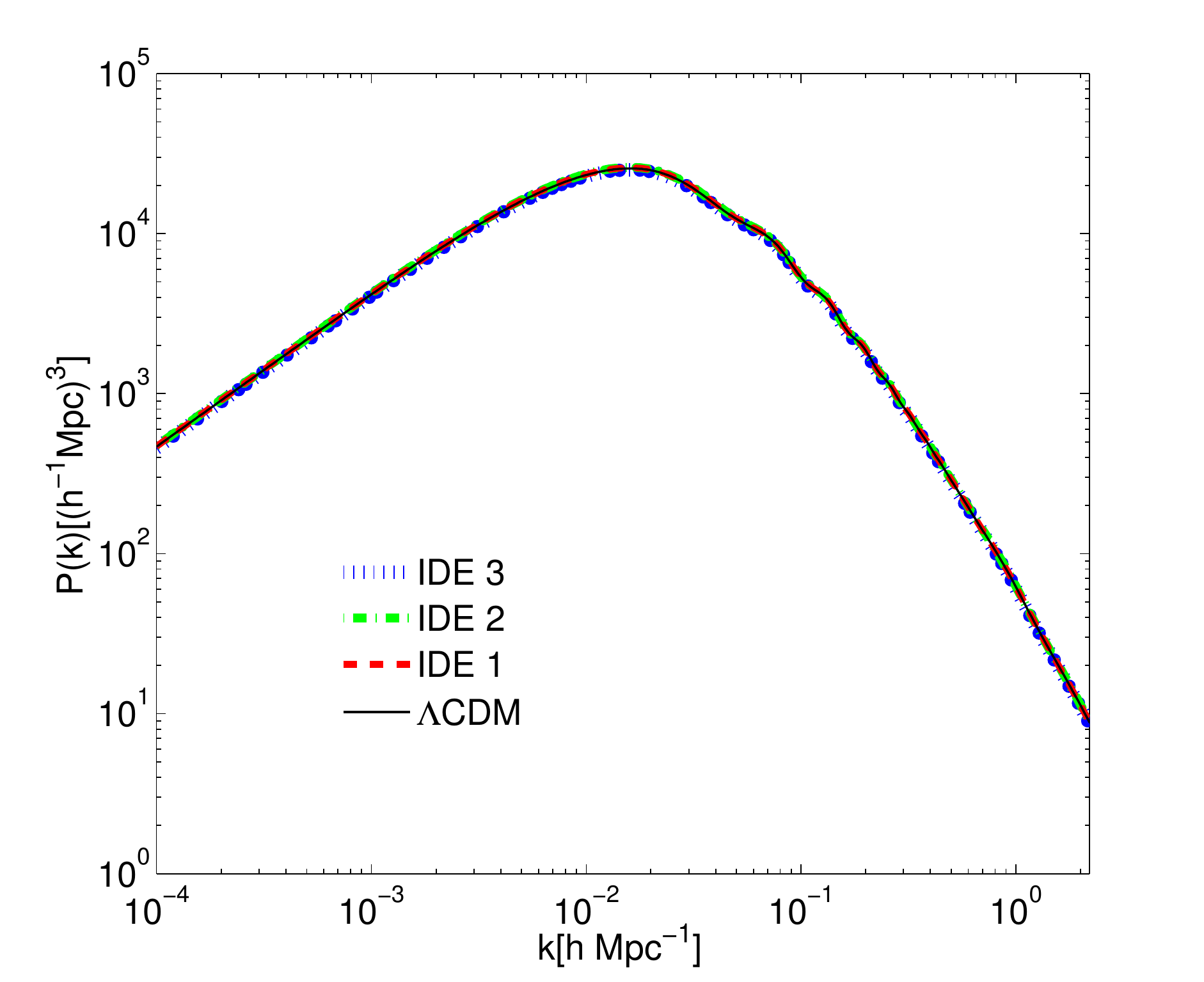}
\caption{\textit{CMB temperature anisotropy (left panel) and the matter
power spectra (right panel) have been shown for three IDE models in compared
to the $\Lambda$CDM model, using the mean values of the free parameters
obtained from the combined analysis CMB $+$ BAO $+$ JLA $+$ RSD $+$ WL $+$
CC $+$ $H_0$.}}
\label{fig:mean}
\end{figure}

\begin{figure}[tbh]
\includegraphics[width=0.32\textwidth]{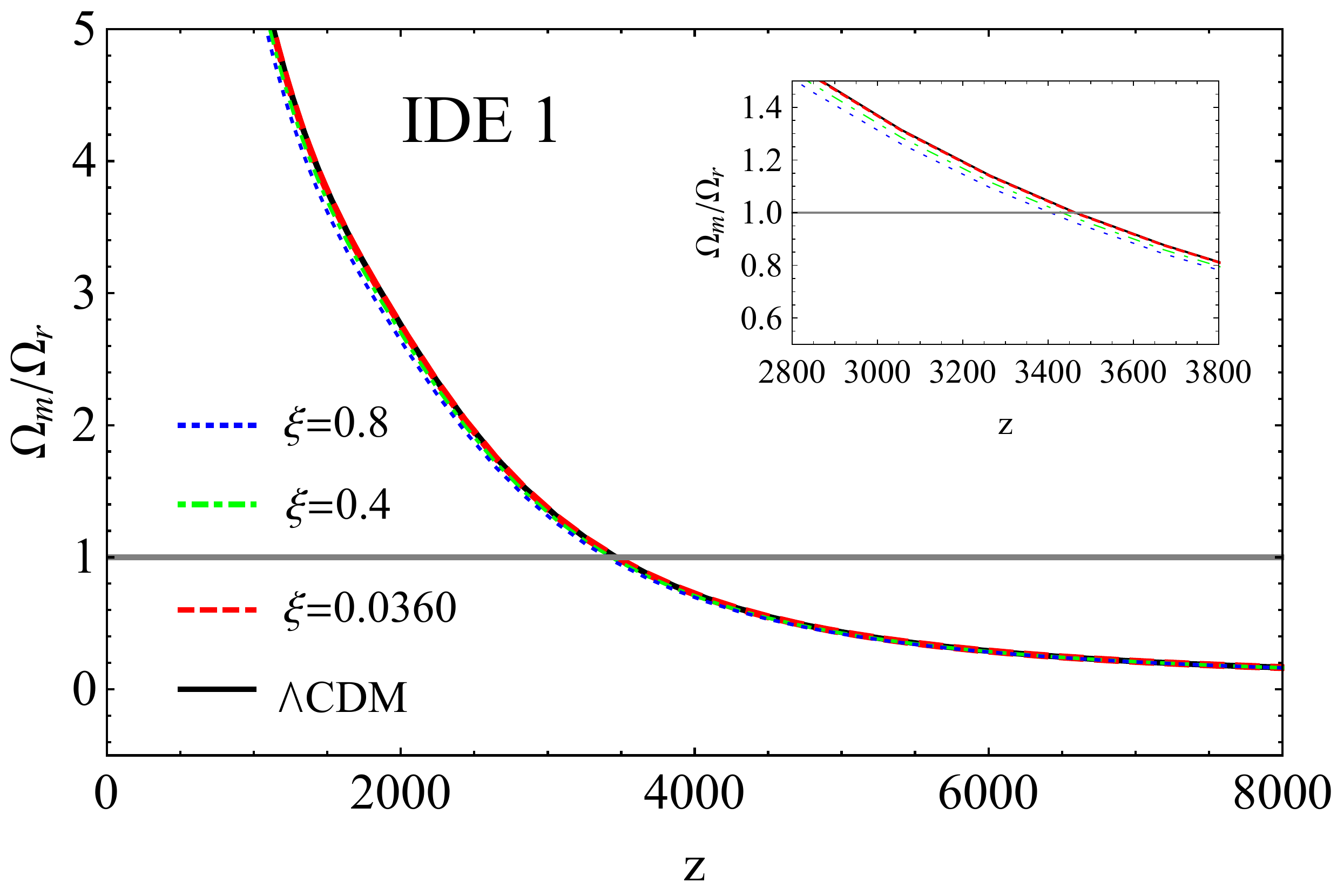} %
\includegraphics[width=0.32\textwidth]{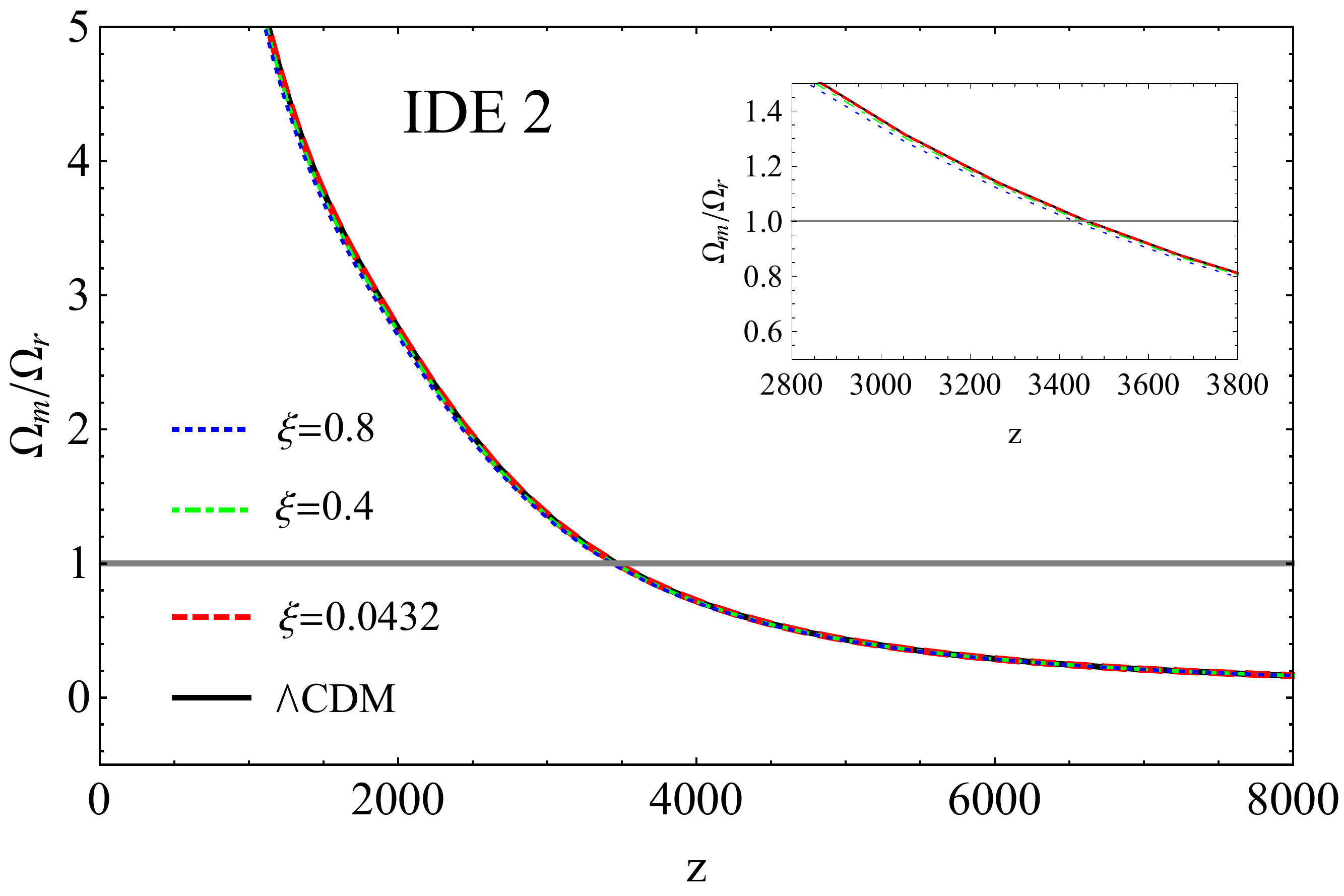} %
\includegraphics[width=0.32\textwidth]{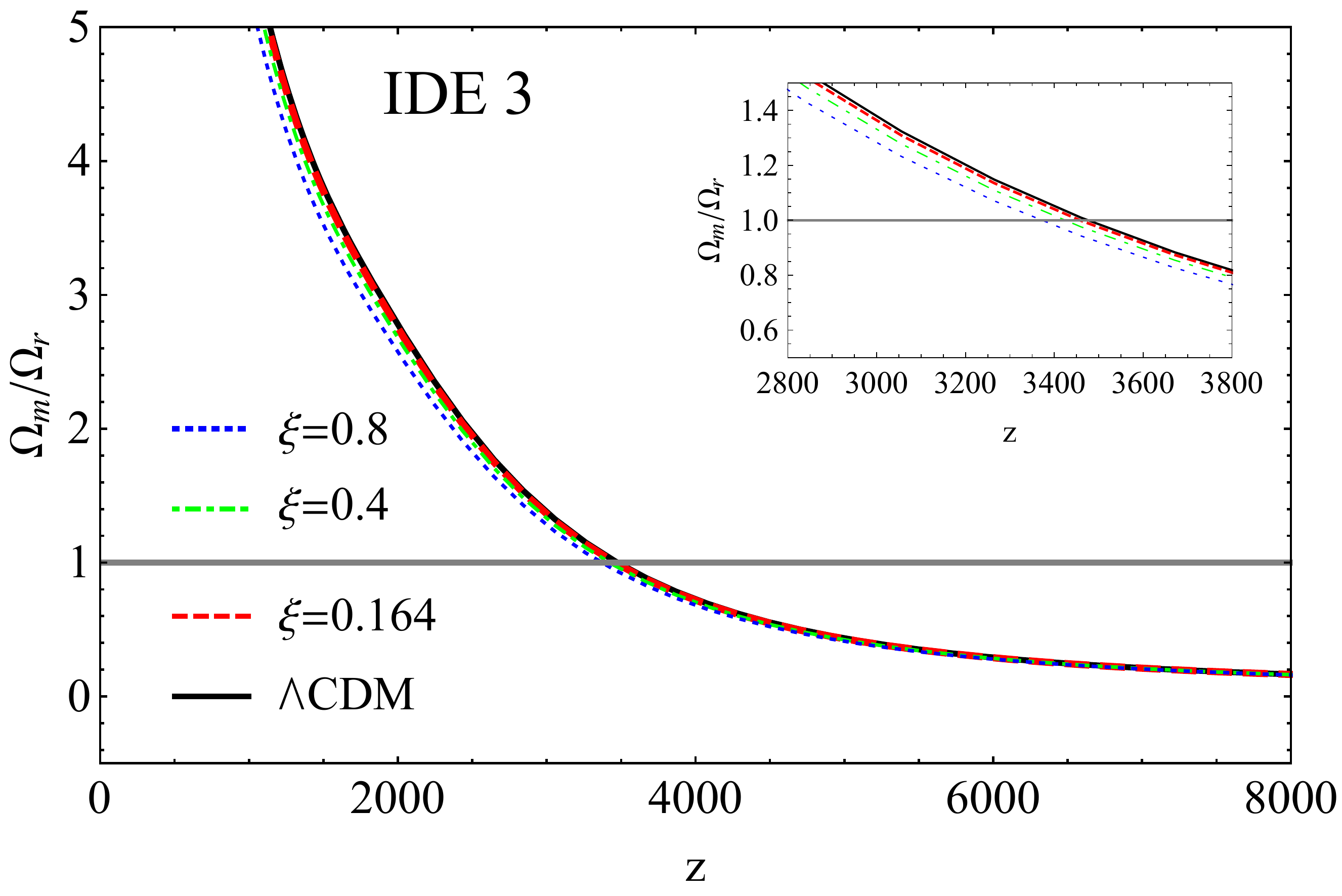}
\caption{\textit{The qualitative evolution of the ratio $\Omega_m/\Omega_r$
(Here $\Omega_m = \Omega_{c}+\Omega_b$) for the three IDE models has been
shown for different values of the coupling parameter $\protect\xi$ and
compared with the $\Lambda$CDM evolution. We note that the values $\protect%
\xi = 0.0360$, $\protect\xi = 0.0432$ and $\protect\xi = 0.164$ are,
respectively, the mean values of the coupling parameters obtained from the
models IDE 1, IDE 2 and IDE 3 using the combined observational analysis CMB $%
+$ BAO $+$ JLA $+$ RSD $+$ WL $+$ CC $+$ $H_0$.}}
\label{fig:ratio}
\end{figure}
\end{center}

\begin{itemize}
\item \textbf{IDE 2}: In Fig. \ref{fig:contour2}, we display the 68.3\% and
95.4\% confidence-level (C.L.) contour plots for different
combinations of the free parameters of this model as well as the one-dimensional posterior distribution for each parameter.
The results for IDE 2 are quite similar to IDE 1. 
The coupling parameter for this model is constrained to be ($%
\xi =0.0433_{-0.0433}^{+0.0062}$ at 68.3\% CL) from the combined analysis,
and we also notice that a zero value of $\xi $ is allowed at the 68.3\% CL.
This means the non-interacting $w_{x}$CDM cosmology is still permitted,
while the observational data always favour $\xi \neq 0$.  In addition, we
find that this interacting scenario allows the equation of state for dark
energy to go over the phantom divide boundary of `$-1$'. The best
fit ($w_{x}=-1.0374$) and the mean value of $w_{x}$ ($%
=-1.0247_{-0.0302}^{+0.0289}$ at 68.3\% CL) are the characteristics of a
phantom dark energy. However, at the 68.3\% CL, the possibility of $w_{x}>-1$
is permitted, at least from the present observational data.
Furthermore, in Figures \ref{fig:cmbplot2} and \ref{fig:Mpower2} we have
plotted the CMB temperature anisotropy spectra and the matter power spectra
for a wide ranges of $w_{x}$ and $\xi ,$ and both these plots indicate that
IDE 2 does not deviate much from the standard $\Lambda $-cosmology. In fact,
we observe that the deviation from the $\Lambda $-cosmology for the strong
coupling, $\xi =0.8,$ is weaker in respect to the deviation for the same
coupling strength observed in IDE 1. A similar argument for $w_{x}$ holds
true as for IDE 1. In Figure \ref{fig:scatterIDE2} we also show the
two-dimensional marginalized posterior distribution for the parameters $%
(w_{x},\xi )$ using the combined analysis mentioned above. The points in
Figure \ref{fig:scatterIDE2} are the samples from the chains of the combined
analysis that have been colored by the values of $H_{0}$. We find similar
behavior as in IDE 1. This means that higher values of $H_{0}$ favor the
phantom regime $w_{x}<-1$ while lower values of $H_{0}$ favor the
quintessence dark energy, i.e. $w_{x}>-1$. The shift from phantom to
quintessence dark energy is displayed as the Hubble parameter values
decrease from higher values. We now return to the estimation of the
Hubble parameter in order to see whether this model could also ease the
tension on $H_{0}$ in a similar fashion to that observed in IDE 1. The
estimated value from our analysis is, $%
H_{0}=68.5099_{-0.9264-1.7640-2.4521}^{+0.8529+2.0520+2.1279}$. One can
clearly see that the inclusion of the coupling shifts the Hubble
parameter towards higher values; however, in comparison to IDE 1, %
the shifting is now slightly higher. Certainly, the tension on $H_{0}$
might be released in a similar fashion. Thus, at the statistical level, this
model resembles IDE 1. Moreover, the estimated value of $\sigma _{8}$ for
this model also matches the $\Lambda $CDM based Planck estimate \cite%
{ref:Planck2015-3}, in the presence of lensing where $\sigma _{8}=0.8149\pm
0.0093$ (Planck TT+lowP+lensing) \cite{ref:Planck2015-3} and $\sigma
_{8}=0.8150\pm 0.0087$ (Planck TT, TE, EE+lowP+lensing) \cite%
{ref:Planck2015-3}. The observational constraints in the presence of the
other data, for instance BAO+JLA+$H_{0},$ return similar fits to $\sigma _{8}
$ \cite{ref:Planck2015-3}. Thus, we see that this interaction model is close
to the $\Lambda $CDM cosmology. \newline

\item \textbf{IDE 3}: In Fig. \ref{fig:contour3}, 
we display the 68.3\% and
95.4\% confidence-level (C.L.) contour plots for different
combinations of the free parameters of this model as well as the one-dimensional posterior distribution for each parameter
The observational constraints on IDE 3
display some different properties to those of IDE 1 and IDE 2. We
find that the coupling parameter $\xi $ is comparatively high ($\xi
=0.1064_{-0.1064}^{+0.0437}$ at 68.3\% CL), unlike in the two other
interaction models (68.3\% CL constraints on the coupling strength are, $\xi
=0.0360_{-0.0360}^{+0.0091}$ for IDE 1 while $\xi =0.0433_{-0.0433}^{+0.0062}
$ for IDE 2), although its zero value is still marginally allowed at the
68.3\% CL. The best fit and the mean values of $w_{x}$ describe a phantom
dark energy. The numerical values of the best fit as well as the
mean values of the dark energy equation of state are, respectively , $%
w_{x}=-1.0134$ and $w_{x}=-1.0275_{-0.0318}^{+0.0228}$ (at 68.3\% CL).  It
is interesting to mention that, at $68.3\%$ CL, the dark-energy equation of
state $w_{x}$ strictly shows phantom behavior. However, at the $95.4\%$
confidence level, $w_{x}$ could still be greater than `$-1$' ($%
w_{x}=-1.0275_{-0.0509}^{+0.0603}$ at 95.4\% CL), that means the
quintessential regime is not excluded at all, at least, with the
present data. Now, following the same trend as in IDE 1 and IDE 2, in
Figures \ref{fig:cmbplot3} and \ref{fig:Mpower3} respectively, we 
show the CMB temperature anisotropy spectra and the matter power spectra for
a wide ranges of $w_{x}$ and the coupling strength $\xi $. From both the
figures, we see that the model shows a clear difference to the $%
\Lambda $-cosmology and hence to the other two interaction models. However,
it is also true that such differences observed in the Figures \ref%
{fig:cmbplot3} and \ref{fig:Mpower3} are not significant enough, although a
non-zero deviation from $\Lambda $-cosmology is clearly presented. The
deviations in other cosmological parameters for this model can also be
compared to IDE 1 and IDE 2. As one can see, a lower value of $\sigma _{8}$ (%
$=0.8051_{-0.0185-0.0396}^{+0.0231+0.0336}$) is favoured for this model
unlike for the other two IDE models where the estimations are, $\sigma _{8}=$
$0.8156_{-0.0137-0.0244}^{+0.0121+0.0246}$ (IDE 1) and $\sigma _{8}=$ $%
0.8166_{-0.0166-0.0280}^{+0.0134+0.0300}$ (IDE 2). This value also reflects
a slight difference from the Planck estimate \cite{ref:Planck2015-3}. Thus
one can see that, according to the observations, this model shows a non-zero
deviation from the $\Lambda $-cosmology with a phantom character within up
to the 68.3\% CL. Now, concerning the tension on $H_{0}$
determinations, we find that IDE 3 may also ease such tension. This might be
clear from the estimation of the Hubble parameter, $%
H_{0}=68.5420_{-0.6763-1.4114-1.9236}^{+0.7817+1.3760+1.6177},$ and by
following similar arguments to those provided for IDE 1 and IDE 2. Finally,
in Figure \ref{fig:scatterIDE3}, we plot the two-dimensional marginalized
posterior distribution for the parameters $(w_{x},\xi )$ as we did for the
models IDE 1 and IDE 2. The observational data are as described above. 
Overall, we find that IDE 3 follows similar trend to IDE 1 and IDE
2, but indeed this interaction model shows differences with respect to the
other two interaction models but such differences are small.
\end{itemize}

\subsection{Comparisons of the IDE models}

\label{subsec-compare}

Let us provide a statistical comparison of the three IDE models. In order to
visualize all three models in a single frame, in Figure \ref{fig:contour-all}
we have provided the contour plots for different combinations of the model
parameters. It is clearly seen that IDE 1 and IDE 2 have considerable
overlap with each other, showing that these two models resemble each other,
while IDE 3 is slightly different which can be seen from the
estimation of the coupling parameter, and also from the behaviour of the
dark energy equation of state which retains its phantom character within
68.3\% CL unlike with other two interaction models, namely IDE 1 and IDE 2.
Nevertheless, they share some common properties. The IDE models all favor a
crossing of the phantom divide line. The mean values and the best fit values
of the dark-energy equation of state all cross the `$-1$' boundary. A
striking feature of all these interaction models is the alleviation of the
tension between the different values of $H_{0}$ deduced from the local \cite%
{Riess:2016jrr} and global measurements \cite{ref:Planck2015-3}. We find
that the allowance of coupling in the dark sector is the main factor that
shifts the the Hubble parameter values toward its local measurement \cite%
{Riess:2016jrr}. We note also that the alleviation of the tension on $H_{0}$
has been found earlier in the context of interacting dark energy \cite%
{Kumar:2016zpg, Kumar:2017dnp, DiValentino:2017iww} with some specific
models. This might be considered as one of the most interesting features of
interacting dark-energy models. \newline

From the analysis of large scale structure it is seen from the evolution of
the matter power spectra (see the right panel of Figure \ref{fig:mean}) or
the CMB temperature anisotropy (see the left panel of Figure \ref{fig:mean})
that the models do not show any remarkable deviation from each other. Within
68.3\% CL, the current interacting models are very close to $\Lambda $CDM
cosmology. Moreover, in Figure \ref{fig:ratio} we have shown the qualitative
evolution of the ratio $\Omega _{m}/\Omega _{r}$ for all interacting models,
and also compared the same evolution with the $\Lambda $-cosmology. We find
that for very small coupling parameter values, the evolution of the quantity 
$\Omega _{m}/\Omega _{r}$ is very close to that of $\Lambda $CDM cosmology. 
However, for larger values of $\xi ~(<1)$, the deviation of course
increases. This is prominent for IDE 3, and then for IDE 1, and after that
for IDE 2 (see the subfigures in Figure \ref{fig:ratio}). But, the
deviations for all three models are not significant enough to draw a
decisive conclusion against the $\Lambda $-cosmology. We note that the
evolution of $\Omega _{m}/\Omega _{r}$ also tells us that IDE 3 is slightly
different from the other two IDE models.  We recall from the temperature
and matter power spectra displayed in Figures \ref{fig:cmbplot3} and \ref%
{fig:Mpower3}, that we noticed similar findings about IDE 3. \newline

\begin{figure}[tbp]
\includegraphics[width=0.45\textwidth]{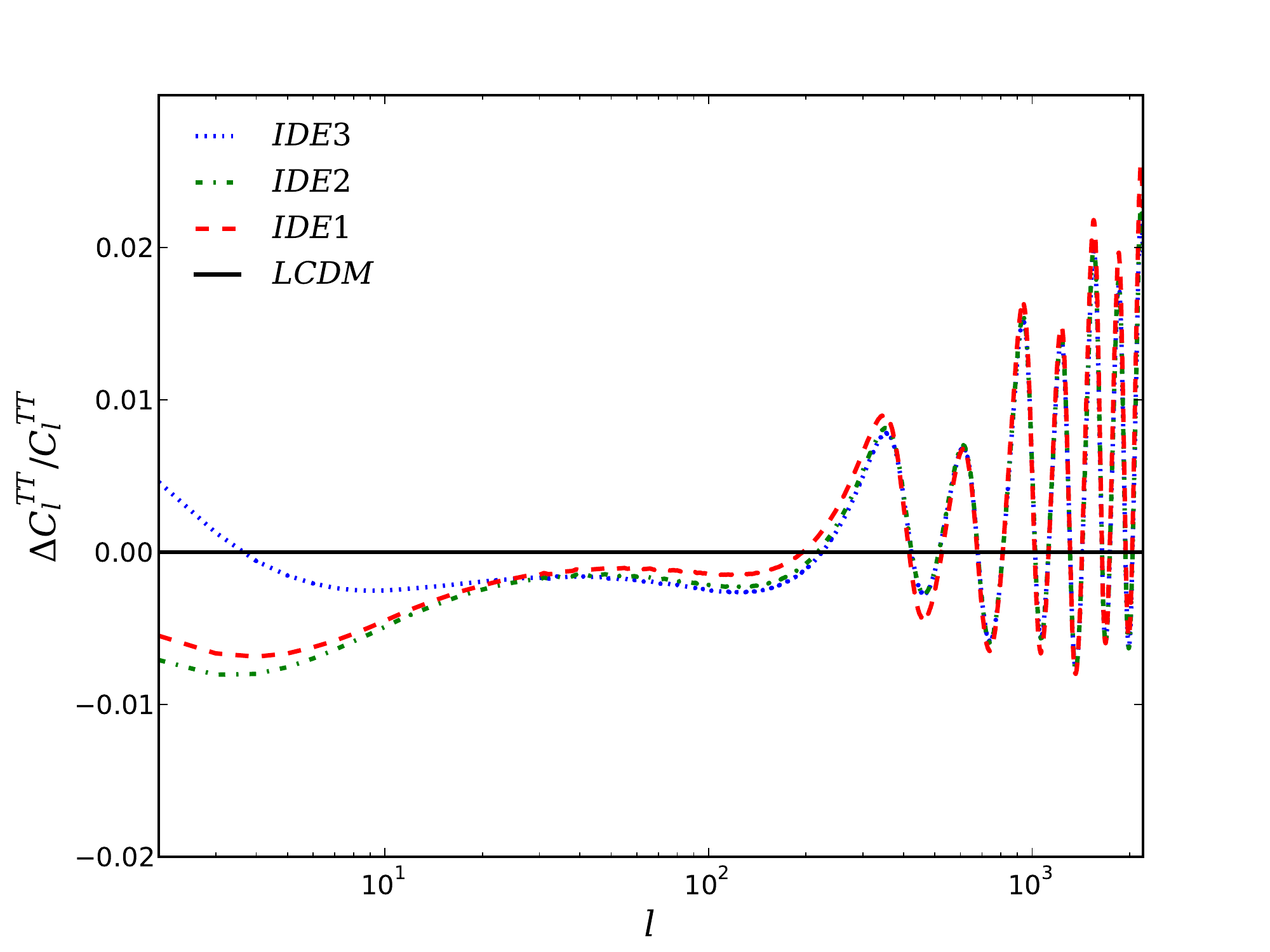} %
\includegraphics[width=0.45\textwidth]{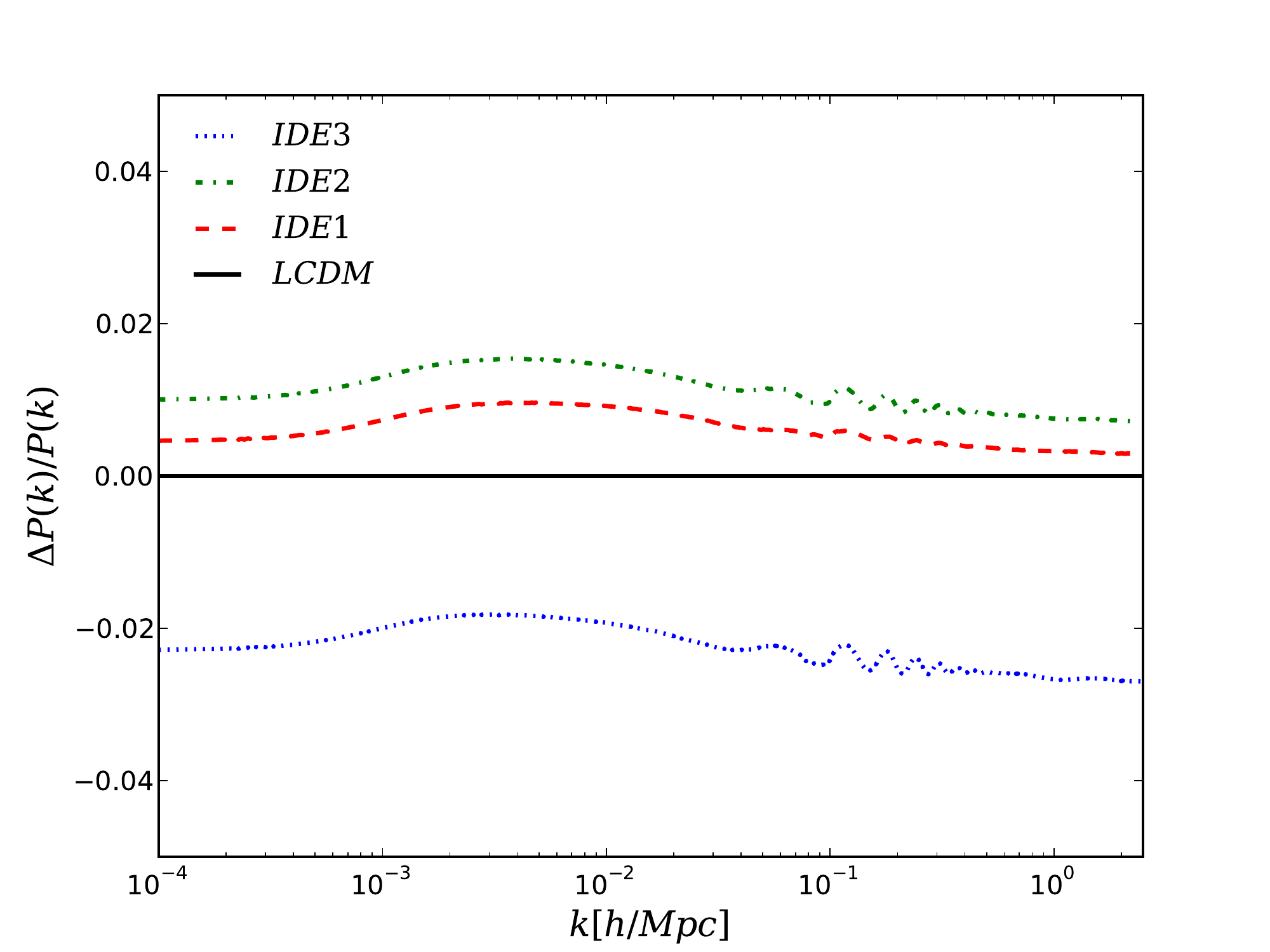}
\caption{\textit{The relative deviations of the IDE models from the $\Lambda$%
-cosmology through the CMB TT and matter power spectra have been shown using
the mean values of the model parameters from the combined analysis CMB $+$
BAO $+$ JLA $+$ RSD $+$ WL $+$ CC $+$ $H_0$. One may notice that the models
are also distinguished from one another. }}
\label{fig-relative-deviation}
\end{figure}

From the temperature anisotropy in the CMB TT spectra and also from the
matter power spectra displayed for all models, the differences between the
different models, as well as from the pure $\Lambda $-cosmology, are not
strong. But, one can clearly show the differences between the models using
the relative deviations of the models with respect to the base $\Lambda $%
-cosmological model. In order to depict the differences between the models,
in Figure \ref{fig-relative-deviation} we have shown the relative deviations
of the models from the pure $\Lambda $-cosmology in terms of the CMB TT
spectra (left-hand panel of Figure \ref{fig-relative-deviation}) and the
matter power spectra (right-hand panel of Figure \ref{fig-relative-deviation}%
) as well. One can clearly see that, the deviations between the models
exist, but such deviations are small. \newline

Finally, we complete our comparisons with a brief remark. In Figures \ref%
{fig:scatterIDE1}, \ref{fig:scatterIDE2} and \ref{fig:scatterIDE3}, we
display the two-dimensional marginalized posterior distribution for the
parameters $(w_{x},\xi )$ using the combined analysis of CMB $+$ BAO $+$ JLA 
$+$ RSD $+$ WL $+$ CC $+$ $H_{0}$. The points in Figures \ref%
{fig:scatterIDE1}, \ref{fig:scatterIDE2} and \ref{fig:scatterIDE3} are the
samples from the chains of the combined analysis that have been colored by
the values of $H_{0}$. We find that for all models the higher values of $%
H_{0}$ favor the phantom regime $w_{x}<-1,$ while the lower values of $H_{0}$
favor a quintessence dark energy, i.e. $w_{x}>-1$. A striking feature
allowed by all the interacting fluid models is that, as the values of $H_{0}$
decrease, a clear shift in the dark-energy behavior, from phantom to
quintessence, is observed, although the dark-energy equation of state still
remains very close to the cosmological constant boundary.\newline

\section{Summary and Conclusions}

\label{sec-summary}

Phenomenological interaction models for the transfer of energies in
cosmological models have been widely investigated in recent years. They
include a wide range of assumed interaction dependences, such as $Q\propto
\rho _{c}$, $Q\propto \rho _{x}$, $Q\propto (\rho _{c}+\rho _{x}),$ and
others. In order to impose observational constraints on these scenarios and
evaluate the stability of the expanding universe models they require, some
specific parametric space needs to be considered. For instance, if the dark
energy equation of state is described by $w_{x}$ and the coupling parameter
of the interaction is $\xi $, then the model is generally tested within two
separate intervals, namely, $w_{x}\geq -1~\&~\xi \geq 0$ or $w_{x}\leq
-1~\&~\xi \leq 0$. So, there exists a discontinuity in the testable range of
the dark energy equation of state. \newline

In this paper we have provided a new technique to test some interacting
models without restriction to any specific subintervals of the parameter
space defining them. We carried out a general analysis of the inhomogeneous
perturbations of a general interaction model linking dark energy and dark
matter. We found that with the introduction of a new factor $(1+w_{x})$ in
the background energy transfer, it is possible to test the whole space of
the equation of state for dark energy with the observational data. We tested
the scenarios using three different interaction models: $Q=3H\xi
(1+w_{x})\rho _{x}$, $Q=3H\xi (1+w_{x})\rho _{c}\rho _{x}/(\rho _{c}+\rho
_{x})$, and $Q=3H\xi (1+w_{x})\rho _{x}^{2}/\rho _{c}$. One can say that the
inclusion of $(1+w_{x})$ into the energy transfer rate $Q$ can be viewed as
a transformation of the coupling parameter as $\xi \rightarrow \bar{\xi}=\xi
(1+w_{x})$. Following this, the models can be viewed in terms of the
transformed coupling parameter $\bar{\xi}$ as $Q=3H\bar{\xi}\rho _{x}$, $Q=3H%
\bar{\xi}\rho _{c}\rho _{x}/(\rho _{c}+\rho _{x})$, and $Q=3H\bar{\xi}\rho
_{x}^{2}/\rho _{c}$. We employed the latest astronomical data from several
independent sources namely the Planck 2015 cosmic microwave background
anisotropy, baryon acoustic oscillation, joint light curves from type Ia
supernovae, redshift space distortions, weak gravitational lensing, cosmic
chronometers together with the best local value of the Hubble constant.
Using the Markov Chain Monte Carlo algorithm we have constrained all three
interaction scenarios. We find that in all these three scenarios,
the observational data favour a non-zero interaction between the dark
sectors. In particular, for the first two IDE models, the observational data
favor an almost zero interaction (the 68.3\% CL constraints are, $\xi
=0.0360_{-0.0360}^{+0.0091}$ for IDE 1 and $\xi =0.0433_{-0.0433}^{+0.0062}$
for IDE 2) while the third one suggests a slightly higher interaction
coupling strength ($\xi =0.1064_{-0.1064}^{+0.0437}$ at 68.3\% CL) in
comparison to the IDE 1 and IDE 2 models.  However, it is clear
that within 68.3\% CL, all interaction models recover the no-interaction
scenario (i.e., $\xi =0$). This means that the observational data allow all
IDE models to converge to the non-interacting $w_{x}$CDM model. %
Furthermore, the observational data also predict that the mean value, as
well as the best fit value, of the dark energy equation of state, $w_{x}$,
both cross the phantom divide line. More precisely, the 68.3\% CL
constraints on the dark energy equation of state for the IDE models are, $%
w_{x}=-1.0230_{-0.0257}^{+0.0329}$ (for IDE 1), $%
w_{x}=-1.0247_{-0.0302}^{+0.0289}$ (for IDE 2), and $%
w_{x}=-1.0275_{-0.0318}^{+0.0228}$ (for IDE 3) As one can see, within 68.3\%
CL, $w_{x}$ is not so far from the cosmological constant boundary `$-1$'.
We\ also observe that all models do not exclude the possibility of $w_{x}>-1$%
. For IDE 1 and IDE 2, $w_{x}>-1$ is allowed in the 68.3\% CL while for IDE
3, 95.4\% CL shows this possibility. Overall, a significant
feature of all these interaction models we find is that, from the analysis
at the background level, none of the our three interaction models can be
distinguished from the $\Lambda $- cosmology. In fact, from the perturbative
analysis, it is also quite difficult to distinguish between the models as
well as distinguish them from the $\Lambda $-cosmology, but of course small
deviations between any two models do exist and all the models also differ
from $\Lambda $-cosmology. Moreover, in all such models, we observe that the
current tension on $H_{0}$ from different data sets can be relieved. This
property of the interacting models could be a general one since some other
recent articles make the same suggestion \cite{Kumar:2017dnp,
DiValentino:2017iww}. Finally, we found that a characteristic feature of all
IDE models is that as the value of Hubble constant decreases, the behavior
of the dark energy equation of state is shifted from phantom to quintessence
type with its equation of state very close to that of a simple cosmological
constant at the present time. \newline

We conclude our analysis with a comparison of the observational constraints
with some of the proposed models, specifically with $Q\propto \rho _{x}$ and 
$Q\propto \rho _{c}\rho _{x}/(\rho _{c}+\rho _{x})$. We note that the
analysis including the cosmological perturbations for the model $Q\propto
\rho _{x}^{2}/\rho _{c}$ has not been performed in past. The differences
between the past and current analyses are that, here we vary the dark energy
equation of state $w_{x}$ within the interval $[-2,0]$, and hence, it is
expected to have slightly different results in compared to the past
analyses. In \cite%
{ref:Yang2014-uc,ref:Yang2014-ux,ref:Yang2014-dh,ref:Yang2016-four}, the
authors performed the analyses for $w_{x}>-1$ that estimated the coupling
parameter for the interaction model $Q=3H\xi \rho _{x}$. In \cite%
{ref:Yang2014-uc}, the authors reported the coupling parameter $\xi $ for
two different sets of the combined analyses that measured $\xi
=0.209_{-0.0403}^{+0.0711}$ at $1\sigma $ confidence-level (for Planck $+$
WMAP9 $+$ SNIa $+$ BAO) and $\xi =0.00372_{-0.00372}^{+0.00768}$ at $1\sigma 
$ confidence-level (for Planck $+$ WMAP9 $+$ SNIa $+$ BAO $+$ RSD) where the
observational data are described in \cite{ref:Yang2014-uc}. Thus, one can
see that the inclusion of RSD into the other data significantly decreases
the coupling strength. Similar analysis can be found in \cite%
{ref:Yang2014-ux, ref:Yang2014-dh, ref:Yang2016-four}. On the other hand, a
recent analysis with $Q\propto \rho _{x}$ where the dark energy equation of
state parameter is constant and allowed to cross the phantom divide line
(i.e. $w_{x}<-1$) \cite{DiValentino:2017iww} shows that within $2\sigma $
confidence-level, the coupling parameter is nonzero ($\xi
=-0.26_{-0.12}^{+0.16}$). Additionally, the interaction model $Q\propto \rho
_{x}$ was tested when the dark energy represents the cosmological constant,
i.e. for $w_{x}=-1$, see the details in \cite{Li:2015vla}. The analysis in 
\cite{Li:2015vla} returned different fits from different observational data,
in particular within $1\sigma $ confidence-level, $\xi
=0.036_{-0.039}^{+0.114}$ (Planck), $\xi =0.020_{-0.053}^{+0.048}$ (Planck $%
+ $ BAO $+$ SNIa), $\xi =-0.026_{-0.053}^{+0.036}$ (Planck $+$ WL $+$ BAO).
In fact, when lensing is added to those data, it is found that the strength
of the interaction decreases for the vacuum interaction scenario, see Table
I of \cite{Li:2015vla} for the details. In the current analysis for $w_{x}$
varying in the interval $[-2,0]$, we obtain similar results to those
obtained in \cite{ref:Yang2014-uc, DiValentino:2017iww, Li:2015vla}. But,
indeed the results should not exactly match with those in refs. \cite%
{ref:Yang2014-uc, DiValentino:2017iww, Li:2015vla} since the astronomical
data do not exactly match ours. We considered next the interaction $Q\propto
\rho _{c}\rho _{x}/(\rho _{c}+\rho _{x})$ constrained in \cite{Li:2013bya}
for $w_{x}>-1$, and the interaction was found to be stable on large scales
provided the coupling parameter was positive. The analysis \cite{Li:2013bya}
found that this nonzero coupling in the dark sector is favoured with $\xi
=0.178_{-0.097}^{+0.081}$ at $1\sigma $ confidence level (Planck $+$ WMAP9 $%
+ $ BAO $+$ SNIa $+$ $H_{0}$). The estimation of the coupling parameter in 
\cite{Li:2013bya} is slightly greater than our estimate for $w_{x}\in
\lbrack -2,0]$. However, we note that the astronomical data in \cite%
{Li:2013bya} and in the current work do not match exactly; thus, the
differences may simply be due to slightly different astronomical data under
consideration. Finally, it might be interesting to make a detailed
comparison with the well known stable interacting dark energy models using
the same astronomical data.

\section*{Acknowledgments}

The authors thank the referees for important comments. The use of Markov
Cahin Monte Carlo package \texttt{CosmoMc} in the analysis of the models is
gratefully acknowledged by the authors. W. Yang's work is supported by the
National Natural Science Foundation of China under Grants No. 11705079 and
No. 11647153. SP was supported by the SERB-NPDF programme (File No.
PDF/2015/000640). J.D. Barrow was supported by the Science and Technology
Facilities Council of the UK (STFC).


\begin{thebibliography}{99}

\bibitem{ref:Planck2015-3} P. A. R. Ade, N. Aghanim, M. Arnaud et al.
[Planck Collaboration], Astron. Astrophys. 594, A13 (2016).

\bibitem{ref:Sofue2001} Y. Sofue and V. Rubin, Ann. Rev. Astron. Astrophys. 
\textbf{39}, 137 (2001).

\bibitem{cop} E. J.~Copeland, M.~Sami and S.~Tsujikawa, Int.\ J.\ Mod.\
Phys.\ D \textbf{15}, 1753 (2006).

\bibitem{at} L. Amendola and S. Tsujikawa, \textit{Dark Energy: Theory and
Observations}, (Cambridge U. P., Cambridge, 2010).

\bibitem{Wetterich1} C. Wetterich, Nucl. Phys. B \textbf{302}, 668 (1988).

\bibitem{Wetterich2} C. Wetterich, Astron. Astrophys. \textbf{301}, 321
(1995).

\bibitem{Lip} S. Z. W. Lip, Phys.Rev. D \textbf{\ 83}, 023528 (2011).

\bibitem{ref:Peebles2010} P. J. E. Peebles, AIP Conf. Proc. \textbf{1241},
175 (2010).

\bibitem{ref:Amendola2000} L. Amendola, Phys. Rev. D \textbf{62}, 043511
(2000).

\bibitem{ref:Amendola2000-2} L. Amendola, Mon. Not. R. Astron. Soc. \textbf{%
312}, 521 (2000).

\bibitem{Billyard:2000bh} A. P.~Billyard and A. A.~Coley, Phys.\ Rev.\ D 
\textbf{61}, 083503 (2000).

\bibitem{ref:Zimdahl2001} W. Zimdahl, D. Pav\'{o}n and L. P. Chimento, Phys. Lett. B \textbf{521}, 133 (2001).

\bibitem{Olivares:2005tb} G.~Olivares, F.~Atrio-Barandela and D.~Pav\'{o}n,
Phys.\ Rev.\ D \textbf{71}, 063523 (2005).

\bibitem{ref:Boehmer2008} C. G. Bohmer, G. Caldera-Cabral, R. Lazkoz and R.
Maartens, Phys. Rev. D \textbf{78}, 023505 (2008).

\bibitem{ref:He2008} J.-H. He and B. Wang, JCAP \textbf{06}, 010 (2008).

\bibitem{Quartin:2008px} M.~Quartin, M. O.~Calvao, S. E.~Joras, R.R.R.~Reis
and I.~Waga, JCAP \textbf{0805}, 007 (2008).

\bibitem{Chimento:2009hj} L. P.~Chimento, Phys.\ Rev.\ D \textbf{81}, 043525
(2010).

\bibitem{ref:Salvatelli2014} V. Salvatelli, N. Said, M. Bruni, A. Melchiorri
and D. Wands, Phys. Rev. Lett. \textbf{113}, 181301 (2014).

\bibitem{Pan:2012ki} S.~Pan, S.~Bhattacharya and S.~Chakraborty, Mon.\ Not.\
Roy.\ Astron.\ Soc.\ \textbf{452}, 3038 (2015).

\bibitem{Nunes:2016dlj} R. C.~Nunes, S.~Pan and E. N.~Saridakis, Phys.\
Rev.\ D \textbf{94}, 023508 (2016).

\bibitem{Kumar:2016zpg} S.~Kumar and R.~C.~Nunes, Phys. Rev. D \textbf{94},
123511 (2016).

\bibitem{Marcondes:2016reb} R. J. F.~Marcondes, R. C. G.~Landim, A.
A.~Costa, B.~Wang and E.~Abdalla, JCAP \textbf{1612}, 009 (2016).

\bibitem{Pan:2016ngu} S.~Pan and G. S.~Sharov, Mon. Not. Roy. Astron. Soc. \textbf{472}, 4736 (2017). 

\bibitem{Mukherjee:2016shl} A.~Mukherjee and N.~Banerjee, Class. Quant.
Grav. \textbf{34}, 035016 (2017).

\bibitem{Sharov:2017iue} G. S.~Sharov, S.~Bhattacharya, S.~Pan, R.~C.~Nunes
and S.~Chakraborty, Mon.\ Not.\ Roy.\ Astron.\ Soc.\ \textbf{466}, 3497
(2017).

\bibitem{Yang:2017yme} W.~Yang, N.~Banerjee and S.~Pan, Phys. Rev. D \textbf{%
95}, 123527 (2017).

\bibitem{ref:Valiviita2008} J. Valiviita, E. Majerotto and R. Maartens, JCAP 
\textbf{07}, 020 (2008).

\bibitem{ref:Majerotto2010} E. Majerotto, J. Valiviita and R. Maartens, Mon.
Not. Roy. Astron. Soc. \textbf{402}, 2344 (2010).

\bibitem{ref:Clemson2012} T. Clemson, K. Koyama, G.-B. Zhao, R. Maartens and
J. Valiviita, Phys. Rev. D \textbf{85}, 043007 (2012).

\bibitem{ref:Yang2014-uc} W. Yang and L. Xu, Phys. Rev. D \textbf{89},
083517 (2014).

\bibitem{ref:Yang2014-ux} W. Yang and L. Xu, JCAP \textbf{08}, 034 (2014).

\bibitem{ref:Yang2014-dh} W. Yang and L. Xu, Phys. Rev. D \textbf{90},
083532 (2014).

\bibitem{ref:Yang2016-four} W. Yang, H. Li, Y. Wu and J. Lu, JCAP \textbf{10}%
, 007 (2016).

\bibitem{Marttens:2016cba} R.~F.~vom Marttens, L.~Casarini, W.~S.~Hip\'{o}%
lito-Ricaldi and W.~Zimdahl, JCAP \textbf{1701}, 050 (2017).

\bibitem{Cai:2017yww} R.~G.~Cai, N.~Tamanini and T.~Yang, 
JCAP \textbf{1705}, 031 (2017).

\bibitem{Barrow:2006hia} J.~D.~Barrow and T.~Clifton, Phys.\ Rev.\ D \textbf{%
73}, 103520 (2006).

\bibitem{ref:RSD-Kaiser1987} N. Kaiser, Mon. Not. Roy. Astron. Soc. \textbf{%
227}, 1 (1987).

\bibitem{ref:RSD-Hamilton1998} A. J. S. Hamilton, Astrophys. Space Sci. Lib. 
\textbf{231}, 185 (1998).

\bibitem{ref:fsigma83-Samushia2012} L. Samushia, W. J. Percival and A.
Raccanelli, Mon. Not. Roy. Astron. Soc. \textbf{420}, 2102 (2012).

\bibitem{ref:fsigma8-DE-Song2009} Y.-S. Song and W. J. Percival, JCAP 
\textbf{10}, 004 (2009).

\bibitem{ref:Bartelmann2011} M. Bartelmann and P. Schneider, Phys. Rept. 
\textbf{340}, 291 (2001).

\bibitem{ref:Heymans2013} C. Heymans et al., Mon. Not. Roy. Astron. Soc. 
\textbf{432}, 2433 (2013).

\bibitem{ref:Heymans2016} M. Asgari, C. Heymans, C. Blake, J.
Harnois-Deraps, P. Schneider and L.V. Waerbeke, Mon.\ Not.\ Roy.\ Astron.\
Soc.\ \textbf{464}, 1676 (2017).

\bibitem{ref:Xu2013-HDE} L. Xu, Phys. Rev. D \textbf{87}, 043525 (2013).

\bibitem{ref:Xu2013-index} L. Xu, Phys. Rev. D \textbf{88}, 084032 (2013).

\bibitem{ref:Xu2013-DGP} L. Xu, JCAP \textbf{02}, 048 (2014).

\bibitem{ref:Yang2013-CASS} W. Yang, L. Xu, Y. Wang and Y. Wu, Phys. Rev. D 
\textbf{89}, 043511 (2014).

\bibitem{ref:Xu2014-vis1} B. Chang and L. Xu, Phys. Rev. D \textbf{90},
027301 (2014).

\bibitem{ref:Xu2014-vis2} B. Chang, J. Lu and L. Xu, Phys. Rev. D \textbf{90}%
, 103528 (2014).

\bibitem{ref:Xu2015-fR} L. Xu, Phys. Rev. D \textbf{91}, 063008 (2015).

\bibitem{ref:Xu2015-MG} L. Xu, Phys. Rev. D \textbf{91}, 103520 (2015).

\bibitem{ref:Xu2015-phiCDM} Y. Chen and L. Xu, Phys. Lett. B \textbf{752},
66 (2016).

\bibitem{ref:zhc2016-TDDE} H. Zhang, E. Li and L. Xu, arXiv:1605.00213.

\bibitem{ref:Ma1995} C. P. Ma and E. Bertschinger, Astrophys. J. \textbf{455}%
, 7 (1995).

\bibitem{ref:Mukhanov1992} V. F. Mukhanov, H. A. Feldman and R.H.
Brandenberger, Phys. Rept. \textbf{215}, 203 (1992).

\bibitem{ref:Malik2009} K. A. Malik and D. Wands, Phys. Rept. \textbf{475},
1 (2009).

\bibitem{ref:Koyama2009} K. Koyama, R. Maartens and Y.-S. Song, JCAP \textbf{%
10}, 017 (2009).

\bibitem{ref:Hu1998} W. Hu, Astrophys. J. \textbf{506}, 485 (1998).

\bibitem{ref:Kodama1984} H. Kodama and M. Sasaki, Prog. Theor. Phys. \textbf{%
78}, 1 (1984).

\bibitem{ref:Gavela2010} M. B. Gavela, L. Lopez Honorez, O. Mena, and S.
Rigolin, JCAP \textbf{11}, 044 (2010).

\bibitem{Li:2013bya} Y.~H.~Li and X.~Zhang, Phys.\ Rev.\ D \textbf{89},
083009 (2014).

\bibitem{ref:Gavela2009} M. B. Gavela, D. Hernandez, L. Lopez Honorez, O.
Mena and S. Rigolin, JCAP \textbf{07}, 034 (2009).

\bibitem{cpl1} M. Chevallier and D. Polarski, Int. J. Mod. Phys. D \textbf{10%
}, 213 (2001).

\bibitem{cpl2} E.~V.~Linder, Phys.\ Rev.\ Lett.\ \textbf{90}, 091301 (2003).

\bibitem{ref:Planck2015-1} R. Adam, P. A. R. Ade, N. Aghanim \textit{et al.}
[Planck Collaboration], Astron. Astrophys. \textbf{594}, A1 (2016).

\bibitem{ref:Planck2015-2} N. Aghanim, M. Arnaud, M. Ashdown \textit{et al.}
[Planck Collaboration], Astron. Astrophys. \textbf{594}, A11 (2016).

\bibitem{Betoule:2014frx} M.~Betoule \textit{et al.} [SDSS Collaboration],
Astron.\ Astrophys.\ \textbf{568}, A22 (2014).

\bibitem{ref:BAO1-Beutler2011} F. Beutler \textit{et al.}, Mon. Not. Roy.
Astron. Soc. \textbf{416}, 3017 (2011).

\bibitem{ref:BAO2-Ross2015} A. J. Ross, L. Samushia, C. Howlett, W. J.
Percival, A. Burden and M. Manera, Mon. Not. Roy. Astron. Soc. \textbf{449},
835 (2015).

\bibitem{ref:BAO3-Gil-Marn2015} H.~Gil-Mar\'{i}n \textit{et al.}, Mon. Not. Roy.
Astron. Soc. \textbf{460}, 4210 (2016).

\bibitem{ref:RSD}
  H.~Gil-Mar\'{i}n \textit{et al.}
  Mon.\ Not.\ Roy.\ Astron.\ Soc.\  {\bf 465}, 1757 (2017).
  


\bibitem{Heymans:2013fya} C.~Heymans \textit{et al.}, Mon.\ Not.\ Roy.\
Astron.\ Soc.\ \textbf{432}, 2433 (2013).

\bibitem{Asgari:2016xuw} M.~Asgari, C.~Heymans, C.~Blake, J.~Harnois-Deraps,
P.~Schneider and L.~Van Waerbeke, Mon.\ Not.\ Roy.\ Astron.\ Soc.\ \textbf{%
464}, 1676 (2017).

\bibitem{Moresco:2016mzx} M.~Moresco \textit{et al.}, JCAP \textbf{1605},
014 (2016).

\bibitem{Riess:2016jrr} A.~G.~Riess \textit{et al.}, Astrophys.\ J.\ \textbf{%
826}, 56 (2016).

\bibitem{gen} J. D. Barrow, Phys. Rev. D \textbf{89}, 064022 (2014).

\bibitem{ref:camb} A. Lewis, A. Challinor and A. Lasenby, Astrophys. J. 
\textbf{538}, 473 (2000).

\bibitem{ref:cosmomc-Lewis2002} A. Lewis and S. Bridle, Phys. Rev. D \textbf{%
66}, 103511 (2002).

\bibitem{Kumar:2017dnp} S.~Kumar and R.~C.~Nunes, 
Phys.\ Rev.\ D \textbf{96}, 103511 (2017). 


\bibitem{DiValentino:2017iww} E.~Di Valentino, A.~Melchiorri and O.~Mena, 
Phys.\ Rev.\ D \textbf{96}, 043503 (2017). 

\bibitem{Li:2013bya} Y.~H.~Li and X.~Zhang, 
Phys.\ Rev.\ D \textbf{89}, 083009 (2014). 

\bibitem{Li:2015vla} Y.~H.~Li, J.~F.~Zhang and X.~Zhang, 
Phys.\ Rev.\ D \textbf{93}, 023002 (2016). 
\end{thebibliography}
\end{document}